\documentclass[pra,reprint,twocolumn,showpacs,superscriptaddress]{revtex4-2}
\usepackage{float}
\usepackage{amsthm}
\usepackage{amssymb}
\usepackage{color}
\usepackage{booktabs}
\usepackage{makecell}
\usepackage{tabularx}
\usepackage{array}
\usepackage{graphicx}
\usepackage{subfigure}
\usepackage{stfloats}
\usepackage{multirow}
\usepackage{mathtools}
\usepackage{dcolumn}
\usepackage{bm}
\usepackage{algorithm2e}
\usepackage[T1]{fontenc}
\usepackage[colorlinks,
            linkcolor=blue,
            anchorcolor=blue,
            citecolor=blue,
            ]{hyperref}
\usepackage{xcolor}
\newtheorem{theorem}{Theorem}
\graphicspath{{data/}}

\begin{document}

\title{Quantifying subspace entanglement with geometric measures}

\author{Xuanran Zhu}
\thanks{These authors contributed equally to this work}
\affiliation{Department of Physics, The Hong Kong University of Science and Technology, Clear Water Bay, Kowloon, Hong Kong, China}

\author{Chao Zhang}
\thanks{These authors contributed equally to this work}
\affiliation{Department of Physics, The Hong Kong University of Science and Technology, Clear Water Bay, Kowloon, Hong Kong, China}

\author{Bei Zeng}
\email{zengb@ust.hk}
\affiliation{Department of Physics, The Hong Kong University of Science and Technology, Clear Water Bay, Kowloon, Hong Kong, China}

\date{\today}

\begin{abstract}
Determining whether a subspace spanned by certain quantum states is entangled and its entanglement dimensionality remains a fundamental challenge in quantum information science. This paper introduces a geometric measure of $r$-bounded rank, $E_r(\mathcal{S})$, for a given subspace $\mathcal{S}$. Derived from the established geometric measure of entanglement, this measure is specifically designed to assess the entanglement within $\mathcal{S}$. It not only serves as a tool for determining the entanglement dimensionality but also illuminates the subspace's capacity to preserve such entanglement. By employing developed non-convex optimization techniques utilized in machine learning area, we can accurately calculate $E_r(\mathcal{S})$ within the manifold optimization framework. Our approach demonstrates notable advantages over existing hierarchical methods, PPT relaxation techniques, and the seesaw strategy, particularly by combining computational efficiency with broad applicability. More importantly, it paves the way for high-dimensional entanglement certification, which is crucial for numerous quantum information tasks. We showcase its effectiveness in validating high-dimensional entangled subspaces in bipartite systems, determining the border rank of multipartite pure states, and identifying genuinely or completely entangled subspaces.
\end{abstract}

\maketitle

\section{Introduction}
Quantum entanglement, a foundational pillar of quantum physics, facilitates the non-local exchange of information amidst entangled particles, regardless of the spatial distance separating them. This fundamental phenomenon underpins transformative applications in realms such as quantum computing \cite{steane1998quantum, preskill2018quantum}, cryptography \cite{gisin2002quantum}, and teleportation \cite{pirandola2015advances}. A critical query within this domain pertains to the determination of whether a given \textit{subspace} is entangled, and if so, discerning its \textit{entanglement dimensionality} \cite{parthasarathy2004maximal, walgate2008generic, demianowicz2022universal, demianowicz2021simple, demianowicz2021simple, bruzda2023rank, eisert2001schmidt, johnston2022complete, demianowicz2019entanglement, wei2003geometric,PRXQuantum.4.020324,nape2021measuring}.

In the bipartite settings, entanglement dimensionality for a pure state can be characterized by Schmidt rank, which can be effectively determined via the Schmidt decomposition \cite{ekert1995entangled}. However, ascertaining the entanglement within a subspace presents a more intricate challenge \cite{buss1999computational}. This process of entanglement certification plays a pivotal role in numerous applications, such as the creation of entangled mixed states \cite{horodecki1997separability}, construction of entanglement witnesses \cite{augusiak2011note, chruscinski2014entanglement}, quantum error correction \cite{gour2007entanglement, huber2020quantum}, and the validation of protocols such as super-dense coding \cite{horodecki2012quantum}.

The complexity of these calculations intensifies in a multipartite context \cite{walter2016multipartite}, even when examining a single multipartite pure state. In contrast to bipartite pure states, which can be represented as matrices, multipartite pure states are depicted by \textit{tensors} \cite{dirac1939new}. While Schmidt decomposition offers a way to determine Schmidt rank, the determination of tensor rank presents a far greater challenge. Specifically, tensor rank, defined as the minimum number of product terms necessary for decomposition, is notoriously difficult to calculate. Its complexity is believed to be at least NP-complete \cite{haastad1990tensor, hillar2013most}.

Meanwhile, the notion of border rank arises \cite{lickteig1984note, landsberg2010ranks}. This concept acknowledges that certain tensors can be approximated \textit{asymptotically} to arbitrary precision using other tensors with smaller tensor ranks—a phenomenon not possible with matrices. A prominent example is the W state in a three-qubit system, which has a tensor rank of 3 but can be approximated by states with a minimum tensor rank of 2, to any desired degree of precision \cite{bruzda2023rank}. In such instances, we denote the border rank of the W state as 2. From a pragmatic perspective, border rank may be more suitable for characterizing entanglement dimensionality in the multipartite scenario. Experimentally, a multipartite pure state with border rank $r$ can always be simulated to any desired precision using states with tensor rank $r$, even if it has a higher tensor rank.

Remarkably, research into tensor or border ranks holds significance not only in the realm of quantum information due to the demand for high-dimensional entanglement in certain applications \cite{erhard2020advances,kong2023high, cozzolino2019high, etcheverry2013quantum} but also in other disciplines, such as algebraic complexity theory \cite{strassen1969gaussian, bini1979n2, bini1980approximate}. Regarding multipartite subspaces, the principal focus rests on completely or genuinely entangled subspaces. The former is characterized as a subspace devoid of any fully product states \cite{bennett1999unextendible}, while the latter contains no product states across any bipartition \cite{demianowicz2018unextendible, agrawal2019genuinely}. The value of completely entangled subspaces stems from their capacity to discriminate pure quantum states locally \cite{walgate2008generic, lovitz2022entangled}. Conversely, genuinely entangled subspaces have proven their worth in the sphere of quantum cryptography \cite{shenoy2019maximally}.

Significant progress has been made in the quest to certify the presence of entanglement within a subspace. One approach involves deriving \textit{lower} bounds on the subspace's entanglement mathematically and then investigating under which conditions these bounds produce a positive value. This line of research has a rich history \cite{cavalcanti2006connecting,ou2007bounds,niset2007tight,song2007bounds,xiang2008bound,akhtarshenas2011concurrence,ma2012improved}, with origins in Ref.~\cite{linden2006entanglement}.
An alternative strategy involves bounding the minimum entanglement using semi-definite programming (SDP), such as the positive partial transpose (PPT) relaxation \cite{demianowicz2019entanglement,zhang2020numerical} and symmetric extension \cite{doherty2004complete}. However, the former mathematical approach may require complex or counterintuitive proofs and may not be broadly applicable. The latter is limited by its computational complexity and is incapable of handling certain scenarios, such as bound entanglement \cite{horodecki1998mixed} and high-dimensional entanglement \cite{cubitt2008dimension}.
Recently, a new method was introduced that employs a hierarchy of linear systems \cite{johnston2022complete}. Although this technique can certify high-dimensional entanglement within the given subspace, there remains room for improvement in its computational efficiency. Additionally, it cannot provide any desired entangled state.

In this work, entanglement dimensionality of a given subspace is characterized by a quantity called \textit{minimal rank}. In the bipartite settings, it is defined as the minimum Schmidt rank of quantum states within the given subspace. And we generalize this definition to the multipartite scenario by replacing Schmidt rank with border rank, based on the consideration mentioned earlier. Here, we also present an intuitive, yet nontrivial approach to determine the minimal rank of a given subspace.

Building upon the geometric measure of entanglement (GME) \cite{demianowicz2021simple, demianowicz2019entanglement,wei2003geometric,blasone2008hierarchies}, which quantifies the geometric ``distance'' between a given pure state $|\psi\rangle$ and fully product states, as follows:
\begin{equation}
\label{GME}
    E_{\text{G}}(|\psi\rangle)=1-\max_{|\varphi_{\text{prod}}\rangle}|\langle\varphi_{\text{prod}}|\psi\rangle|^2,
\end{equation}
we generalize this concept to the geometric measure of $r$-bounded rank for a given subspace $\mathcal{S}$, dentoed by $E_r(\mathcal{S})$, as the geometric ``distance'' between $\mathcal{S}$ and $\sigma_{r-1}$, where $\sigma_{r-1}$ represents the states with at most tensor rank $r-1$. The minimal rank $r(\mathcal{S})$ of a subspace $\mathcal{S}$ can now be determined by finding the largest value of $r$ such that $E_r(\mathcal{S})>0$ [if $E_2=0$ then $r(\mathcal{S})=1$]. Through further analysis, we find that this kind of geometric measure not only provides a criterion for certifying the entanglement dimensionality within a given subspace (from zero or non-zero), but also furnishes insights into the ability of the subspace to maintain such entanglement (indicated by the magnitude).

By employing the gradient back-propagation technique \cite{pytorch} from the field of machine learning, we are able to calculate this geometric measure through gradient descent with notable efficiency, which traditional convex optimization methods like SDP cannot handle. The core concept involves transforming the optimization problem defined on the complicated set $\sigma_r$ into an optimization over a product manifold composed of several simpler manifolds. We then apply the concept of \textit{trivialization} \cite{lezcanocasado2019trivializations}, established in the domain of manifold optimization \cite{hu2020brief, absil2008optimization}, to address the original problem within an unconstrained Euclidean space.

We demonstrate its effectiveness in validating high-dimensional entangled subspaces in bipartite systems, determining the border rank of multipartite states, and identifying genuinely or completely entangled subspaces. The results reveal that this method has some attractive advantages compared to other known approaches, especially as it combines computational efficiency and broad applicability. More importantly, it paves the way for high-dimensional entanglement certification, which is crucial for numerous quantum information tasks. A preview of the main results is presented in Table.~\ref{preview}.

The structure of this paper is as follows: In Sec.~\ref{pre}, we introduce some fundamental definitions related to bipartite and multipartite entanglement. In Sec.~\ref{method}, we propose the geometric measure of $r$-bounded rank and shed light on the connections among pure states, mixed states, and subspaces. We also briefly explain the key idea about the numerical calculation proposed in the manifold optimization framework, with the developed non-convex optimization techniques. In Sec.~\ref{results}, we illustrate the application of this type of geometric measure in various situations and also provide comparisons with other methods. Conclusions and outlooks are summarized and discussed in Sec.~\ref{conclusions}.

\begin{table}[H]
    \centering
    \caption{\label{preview} Comparison between different methods for detecting entanglement: Hier (hierarchical method \cite{johnston2022complete}), PPT (postive partial transpose relaxation \cite{demianowicz2019entanglement,zhang2020numerical}) and the method we propose based on GD (gradient descent approach). The presence of a checkmark ($\checkmark$) indicates the method's capability to address the specified scenario, while a crossmark ($\times$) implies its limitation. Entanglement is denoted as ENT for simplicity.}
    \begin{tabular}{|c|c|c|c|c|}
        \hline
        \multicolumn{2}{|c|}{\textbf{Comparison Criterion}} & Hier & PPT & \textbf{GD} \\
         \hline
        \multirow{3}*{Bipartite} & low-dimensional ENT & $\checkmark$ & $\checkmark$ & $\checkmark$ \\
        \cline{2-5}
         & bound ENT & $\checkmark$ & $\times$ & $\checkmark$ \\
         \cline{2-5}
         & high-dimensional ENT & $\checkmark$ & $\times$ & $\checkmark$ \\
        \hline
          \multirow{3}*{Multipartite} & border rank of pure state & $\times$ & $\times$ & $\checkmark$ \\
         \cline{2-5}
          & complete ENT & $\checkmark$ & $\checkmark$ &$\checkmark$ \\
          \cline{2-5}
          & genuine ENT & $\checkmark$ & $\checkmark$ &$\checkmark$ \\
          \hline
          \multicolumn{2}{|c|}{Quantify the robustness?} & $\times$ & $\checkmark$ & $\checkmark$ \\
          \hline
          \multicolumn{2}{|c|}{Time efficiency} & $\checkmark$ &$\checkmark$ $\checkmark$  & $\checkmark$ $\checkmark$ $\checkmark$ \\
        \hline
    \end{tabular}
\end{table}

\section{Preliminaries\label{pre}}

\textit{Notation}. In the paper, $\vert \psi\rangle$, $\vert \phi\rangle$, state, etc., all refer to normalized quantum states unless otherwise specified.
\subsection{Bipartite entanglement}
\textbf{Schmidt decomposition}.
Let us begin with the simplest bipartite case and consider a Hilbert space $\mathcal{H}=\mathcal{H}_1\otimes\mathcal{H}_2$. A state $\vert \psi \rangle \in \mathcal{H}$ is \textit{separable} iff $\vert \psi \rangle=\vert \phi_1 \rangle \otimes \vert \phi_2 \rangle$ for some pure states $\vert \phi_i \rangle \in \mathcal{H}_i$; otherwise it is called \textit{entangled}. A useful tool for the characterization of bipartite entanglement is the \textit{Schmidt decomposition}: any $\vert \psi \rangle \in \mathcal{H}$ can be written as
\begin{equation}
\label{schmidt decomposition}
	|\psi\rangle=\sum_{i=1}^r \lambda_i\left|e_i\right\rangle\left|f_i\right\rangle,
\end{equation}
where the coefficients $\lambda_i$ are positive numbers that can be ordered as $\lambda_1 \geq \lambda_2 \geq \ldots \geq \lambda_r >0$, whose squares sum up to one, $\{\vert e_i\rangle\}$ and $\{\vert f_i\rangle\}$ are orthonormal bases of the local Hilbert spaces and $r \leq \min(d_1,d_2)$ is called the \textit{Schmidt rank} of $\vert \psi \rangle$ representing the minimum number of terms for decomposing a state. It is equivalent to the singular value decomposition (SVD) of the corresponding coefficient matrix. Therefore, in this paper, we will use matrix rank and Schmidt rank interchangeably, as they refer to the same concept.

\textbf{Minimal rank of bipartite subspace}. Given a subspace $\mathcal{S}$ spanned by a set of \textit{bipartite} states $\{ \vert \psi_1\rangle, \vert \psi_2\rangle,\dots,\vert \psi_m\rangle\}$, \textit{minimal rank} of  $\mathcal{S}$ is defined as the following:
\[
\operatorname{r}(\mathcal{S})=\min_{\vert\psi\rangle \in \mathcal{S}}R_S(\vert\psi\rangle),
\]
where $R_S$ represents the schmidt rank. $\vert \psi \rangle \in \mathcal{S}$ means it can be derived by a combination of the states within $\mathcal{S}$, i.e.,
$
\vert \psi \rangle=\sum_{i=1}^{m}c_i\vert \psi_i\rangle,
$
where $c_i$'s are some complex coefficients constrained to the normalization condition. If the minimal rank of a given subspace is larger than one, we say that the subspace is entangled.
\subsection{Multipartite entanglement}
\textbf{Tensor decomposition}. For a general multipartite pure state $\vert \psi \rangle \in \mathcal{H}_1 \otimes\mathcal{H}_2\otimes\cdots\otimes\mathcal{H}_n$, \textit{tensor rank} is defined as the minimum number $r$ such that there exist states $\vert \phi_i^{(k)}\rangle \in \mathcal{H}_k (1\leq i\leq r, 1\leq k\leq n)$ satisfying
\begin{equation}
    \vert \psi \rangle = \sum_{i=1}^r \lambda_i \vert \phi_i^{(1)}\rangle \otimes \vert \phi_i^{(2)}\rangle \otimes \cdots \otimes \vert \phi_i^{(n)}\rangle, \label{eq:tensor-decomp}
\end{equation}
where the coefficients $\lambda_i$ are positive but the different product terms may not be orthonormal to each other.

Tensor rank is a legitimate entanglement measure generalized from the Schmidt rank \cite{carteret2000multipartite, eisert2001schmidt} since it is a strictly nonincreasing quantity under local operations (even stochastic) \cite{chen2010tensor}. It provides a method of detecting multipartite entanglement: a multipartite pure state is entangled if and only if its tensor rank is larger than one. 

\textit{Border rank} is defined as the minimum number $r$ of product terms that are sufficient to approximate the given tensor \textit{asymptotically} with arbitrarily small error (in this paper, we choose the geometric measure).

A famous example of the gap between tensor rank and border rank is that the tensor rank of the W state in three-qubit is 3 while its border rank is 2 since
\begin{equation}
\label{W state}
    \left|W\right\rangle=\frac{1}{\sqrt{3}}\left[\lim _{t \rightarrow 0} \frac{1}{t}\left(\left(|0\rangle+t |1\rangle\right)^{\otimes 3}-|0\rangle^{\otimes 3}\right) \right].
\end{equation}
For any $t \neq 0$, the tensor on the right-hand side has rank 2 and in the limit $t \rightarrow 0$, it becomes W state. For brevity, in the following, we will denote the tensor and border rank as $R$ and $\underline{R}$ separately.

The gap between these two different ranks appears due to the non-existence of a \textit{best rank-r approximation} for an arbitrary tensor $A$ \cite{de2008tensor}, implying $R(A)> \underline{R}(A)$. However, there exist two notable exceptions where $R=\underline{R}$: the cases $n=2$ (the given tensors are matrices) and $r=1$ (approximation by rank-1 tensors). The former implies $R(A)=\underline{R}(A)$ for a matrix $A$ while the latter means that $\underline{R}(A)=1$ is equivalent to with $R(A)=1$.

\textbf{Minimal rank of multipartite subspace}. Given a subspace $\mathcal{S}$ spanned by a set of \textit{multipartite} states $\{ \vert \psi_1\rangle, \vert \psi_2\rangle,\dots,\vert \psi_m\rangle\}$,  \textit{minimal rank} of $\mathcal{S}$ is defined as the following:
\[
\operatorname{r}(\mathcal{S})=\min_{\vert\psi\rangle \in \mathcal{S}}\underline{R}(\vert\psi\rangle).
\]
In this work, we choose the border rank as the definition of minimal rank in the multipartite settings because, as mentioned before, experimentally, a multipartite pure state of border rank $r$ can always be simulated to any desired precision using those states of tensor rank $r$ even if it has higher tensor rank. This definition is also consistent across both bipartite and multipartite cases, as the Schmidt rank is equivalent to the border rank for bipartite states.

\textbf{Completely entangled subspace}. Consider a subspace $\mathcal{S} \subset \mathcal{H}_1 \otimes\mathcal{H}_2\otimes\cdots\otimes\mathcal{H}_n$, we call $\mathcal{S}$ is \textit{completely entangled} iff it contains no \textit{fully product} state. Due to the existence of best rank-1 approximation as mentioned above, we know it is equivalent to the detection of subspaces with $r(\mathcal{S}) > 1$.

\textbf{Genuinely entangled subspace}. A subspace $\mathcal{S}\subset \mathcal{H}_1 \otimes\mathcal{H}_2\otimes\cdots\otimes\mathcal{H}_n$ is \textit{genuinely entangled} iff it contains no product state for any \textit{bipartition} $K \vert K^c$, where $K$ is a subset of $\{1,2,\dots,n\}$ and $K^c$ denotes the complementary set. For example, for a tripartite state $\vert \psi \rangle \in \mathcal{H}_A \otimes \mathcal{H}_B \otimes \mathcal{H}_C$, there are three possible bipartition $AB\vert C$, $AC\vert B$ and $BC\vert A$. The genuine entanglement of a subspace is equivalent to $r(\mathcal{S})>1$ for any bipartition $K\vert K^c$.

\section{Methodology}\label{method}

\subsection{Geometric measure for entangled subspace}
Geometric measure for a pure state $\vert \psi \rangle $ is defined through the following general formula
\[
E(|\psi\rangle)=1-\max _{|\varphi\rangle \in V}|\langle\varphi \vert \psi\rangle|^2,
\]
where $V$ is chosen according to the specific purpose.

In order to quantify entanglement dimensionality via minimal rank, we  generalize the GME defined in Eq.~(\ref{GME}) to the geometric measure of $r$-bounded rank
\begin{equation}
\label{gm-r}
    E_r(|\psi\rangle)=1-\max _{|\varphi\rangle \in \sigma_{r-1}}|\langle\varphi \vert \psi\rangle|^2.
\end{equation}
Here, $\sigma_{r-1}$ is the states with at most tensor rank $r-1$. Obviously, it can be reduced into the form of geometric measure of entanglement when $r=2$, i.e., $E_G(|\psi\rangle)=E_2(|\psi\rangle)$.

For a bipartite state $|\psi\rangle$ it has been shown that \cite{demianowicz2021simple}
\[
E_r(|\psi\rangle)=1-\left(\lambda_1^2+\lambda_2^2+\cdots+\lambda_{r-1}^2\right),
\]
where $\lambda_i$ is the coefficients after the Schmidt decomposition [see Eq.~(\ref{schmidt decomposition})].  As for a multipartite state $\vert \psi \rangle$ of border rank $r$, $r$ is the largest value that makes $E_r(\vert \psi \rangle)>0$. Thus, this measure can serve as a tool for certifying the border rank of a given state.

We can generalize this geometric measure, defined for a pure state, to a subspace $\mathcal{S}$ as the following:
\begin{equation}
\label{gm-r-subspace}
\begin{aligned}
 E_r(\mathcal{S}) &= \min _{|\psi\rangle \in \mathcal{S}} E_r(|\psi\rangle) \\
& =\min _{|\psi\rangle \in \mathcal{S}}(1-\max _{\vert \varphi \rangle \in \sigma_{r-1}}|\langle\varphi \vert \psi\rangle|^2) \\
& =1-\max _{\vert \varphi \rangle \in \sigma_{r-1}} \max _{|\psi\rangle \in \mathcal{S}}|\langle\varphi \vert \psi\rangle|^2\\
& =1-\max _{\vert \varphi \rangle \in \sigma_{r-1}}\left\langle\varphi\left|\mathcal{P}_{\mathcal{S}}\right| \varphi\right\rangle \\
& =\min _{\vert \varphi \rangle  \in \sigma_{r-1}}\left\langle\varphi\left|\mathcal{P}_{\mathcal{S}}^{\perp}\right| \varphi\right\rangle,
\end{aligned}
\end{equation}
where $\mathcal{P}_{\mathcal{S}}$ projects onto $\mathcal{S}$ and $\mathcal{P}_{\mathcal{S}}^{\perp}$ onto the orthogonal complementary space $\mathcal{S}^{\perp}$. The most crucial step is the transition from the third line to the fourth line follows from the fact that for a given $\vert \varphi \rangle$, the vector from $\mathcal{S}$ maximizing the quantity will be the projection of $\vert \varphi \rangle$ onto $\mathcal{S}$. Similarly, for a subspace $\mathcal{S}$ of minimal rank $r$, $r$ is the largest value that makes $E_r(\mathcal{S})>0$.

Besides the certification of entanglement dimensionality, $E_r$ also characterizes the \textit{robustness} of entangled subspaces due to the following theorem (some similar consideration appears in \cite{cavalcanti2006connecting}):
\begin{theorem}
    (\textit{Robustness of entangled subspace}). Given a subspace $\mathcal{S} \subset \mathcal{H}_1\otimes \mathcal{H}_2$ with $r(\mathcal{S}) \geq r$, for any perturbation $U=e^{-iH}$ with the trace norm of $H$ less than $E_{r}(\mathcal{S})^{\frac{1}{2}}$, the subspace after the perturbation $\mathcal{S}^{\prime}$ will maintain the minimal rank $r(\mathcal{S}^{\prime}) \geq r$.
\end{theorem}
\begin{proof}
Suppose $\mathcal{S}$ is spanned by $\{ \phi_1,\phi_2,\dots, \phi_m\}$, the subspace after some perturbation $U$ is defined as $\mathcal{S}^{\prime}$, spanned by $\{ U\phi_1,U\phi_2,\dots, U\phi_m\}$. It is easy to prove that
\[
\mathcal{P}_{\mathcal{S}^{\prime}}^{\perp}=U\mathcal{P}_{\mathcal{S}}^{\perp}U^{\dagger}.
\]
For convenience, we denote the square root of $\left\langle x, \mathcal{P}_{\mathcal{S}}^{\perp} x \right\rangle$ as $F_{\mathcal{S}}(x)$. Obviously, there exist the following inequalities:
\[
\begin{gathered}
    F_{\mathcal{S}}(x)+F_{\mathcal{S}}(y)\geq F_{\mathcal{S}}(x+y) \geq F_{\mathcal{S}}(x)-F_{\mathcal{S}}(y),\\
    F_{\mathcal{S}}(\sum_i a_i x_i)\leq \sum_i|a_i|\cdot F_{\mathcal{S}}(x_i),
\end{gathered}
\]
where $x, y, x_i$ can be unnormalized states and $a_i$'s are complex numbers. Thus, the square root of the geometric measure of $r$-bounded rank for $\mathcal{S}^{\prime}$

\[
\begin{aligned}
	E_{r}(\mathcal{S}^{\prime})^{\frac{1}{2}} =&\min _{\vert \varphi \rangle \in \sigma_{r-1}}(\langle\varphi|\mathcal{P}_{\mathcal{S}^{\prime}}^{\perp}}| \varphi\rangle)^{\frac{1}{2}\\
	=& \min _{\vert \varphi \rangle \in \sigma_{r-1}}(\langle\varphi| U \mathcal{P}_{\mathcal{S}}^{\perp}U^{\dagger}| \varphi\rangle)^{\frac{1}{2}}\\
	\geq & \min _{\vert \varphi \rangle \in \sigma_{r-1}}[F_{\mathcal{S}}(\vert \varphi \rangle)-F_{\mathcal{S}}((I-U^{\dagger})\vert \varphi \rangle)]\\
	= & E_{r}(\mathcal{S})^{\frac{1}{2}}-\max _{\vert\varphi\rangle \in \sigma_{r-1}}F_{\mathcal{S}}((I-U^{\dagger})\vert \varphi \rangle).
\end{aligned}
\]
Here, $I-U^{\dagger}$ can be diagonalized into $\sum_i(1-e^{i h_i})\vert u_i\rangle\langle u_i \vert$, where $h_i$ is the eigenvalue of Hamiltonian $H$. Then
\[
\begin{aligned}
	\max _{\vert\varphi\rangle \in \sigma_{r-1}} & F_{\mathcal{S}}((I-U^{\dagger})\vert \varphi \rangle) \\= &\max _{\vert\varphi\rangle \in \sigma_{r-1}}F_{\mathcal{S}}(\sum_i(1-e^{i h_i})\vert u_i\rangle\langle u_i \vert \varphi \rangle) \\
	\leq & \max _{\vert\varphi\rangle \in \sigma_{r-1}}\sum_i\vert1-e^{i h_i}\vert \cdot F_{\mathcal{S}}(\vert u_i\rangle\langle u_i \vert \varphi \rangle) \\
	\leq & \sum_i\vert1-e^{i h_i}\vert \leq \sum_i \vert h_i \vert = \|H\|_{tr}.
\end{aligned}
\]
Therefore, if the trace norm of Hamiltonian $H$ is less than $E_{r}(\mathcal{S})^{\frac{1}{2}}$, $E_{r}(\mathcal{S}^{\prime})$ will be nonzero, i.e., $r(\mathcal{S}^{\prime}) \geq r$.
\end{proof}

\begin{figure}[H]
    \centering
    \includegraphics[width=\linewidth]{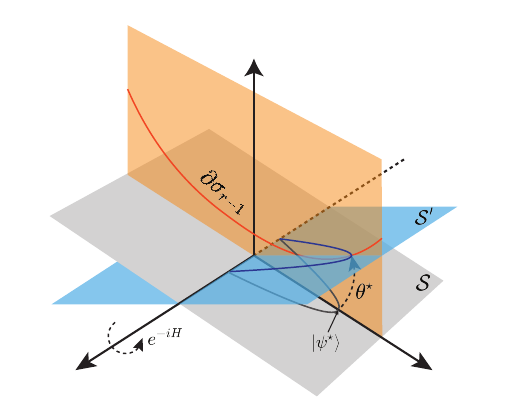}
    \caption{Geometric interpretation of the robustness of entangled subspace. The gray line represents the quantum states within the subspace $\mathcal{S}$, and $|\psi^{\star}\rangle$ is the closest state to the $\sigma_{r-1}$ under the geometric measure. The blue line represents the quantum states within the subspace $\mathcal{S}^{\prime}$ after the perturbation. The orange line denotes the boundary of $\sigma_{r-1}$. }
    \label{fig:subspace_rotation}
\end{figure}

A simple geometric interpretation can be made for this theorem, as shown in Fig.~\ref{fig:subspace_rotation}: notice that $E_r(\mathcal{S})^{\frac{1}{2}}$ characterizes the sine value of the \textit{smallest angle} $\theta^{\star}$ between the states in $\sigma_{r-1}$ and states in the given subspace $\mathcal{S}$, see Eq.~(\ref{gm-r}). A perturbation $U=e^{-iH}$ can be considered as a rotation operator in high-dimensional space, the trace norm of $H$ indicates its rotation angle. In order to maintain its entanglement dimensionality, we need to limit the rotation angle $\theta \leq \theta^{\star}$.

The definition can also be extended into mixed states as follows,

\[
E_r(\rho)=\min_{\{ p_i,\vert \psi_i \rangle\}}\sum_i p_i E_r(\vert \psi_i \rangle),
\]
where the minimum is computed over all pure state ensembles, i.e., $\rho = \sum_i p_i \vert \psi\rangle \langle\psi \vert$.
Obviously, $\rho$ is separable iff $E_2(\rho) =0$, i.e., it can be decomposed into a product state ensemble. It is known that $E_2(\rho) \geq E_2[\text{supp}(\rho)]$, where $\text{supp}(\rho)$ is the support space of the density matrix $\rho$ \cite{branciard2010evaluation}. This result can be generalized to the high-dimensional entangled cases easily, i.e., $E_r(\rho) \geq E_r[\text{supp}(\rho)]$ for $r>2$.

Based on Theorem 1 and the above definition for mixed states, we can further conclude that quantum states supported on the robust entangled subspaces are also robust:
\begin{theorem}
    (\textit{Formation of robust entangled states}). Given a entangled state $\rho \in \mathcal{H}_1 \otimes \mathcal{H}_2$ with the support space $\text{supp}(\rho)$, $E_2[\text{supp}(\rho)]>0$. For any perturbation $U=e^{-iH}$ with the trace norm of $H$ less than $E_2[\text{supp}(\rho)]^{\frac{1}{2}}$, the mixed state after the perturbation $\rho^{\prime}=U\rho U^{\dagger}$ will still be entangled.
\end{theorem}
\begin{proof}
	Suppose $\rho = \sum_i p_i \vert \psi_i \rangle \langle \psi_i \vert$, the corresponding support space $\text{supp}(\rho)$ is spanned by $\{\vert \psi_i \rangle \}$. After the perturbation, it becomes $\rho^{\prime}= \sum_ip_i U\vert \psi_i \rangle \langle \psi_i \vert U^{\dagger}$, the support space becomes $\text{supp}(\rho^{\prime})$ spanned by $\{ U\vert \psi_i \rangle\}$. According to Theorem 1, if $\|H\|_{tr} < E_2[\text{supp}(\rho)]^{\frac{1}{2}}$, then $E_2(\rho^{\prime}) \geq E_2[\text{supp}(\rho^{\prime})]>0$.
\end{proof}
This theorem also offer a feasible way to construct robust entangled states from the robust entangled subspace, which can also be generalized to high-dimensional entanglement with $r>2$.

In conclusion, this kind of geometric measure serves a dual purpose: it provides a criterion for certifying the entanglement dimensionality within a given subspace, quantified by the minimal rank, and offers insights into the robustness of the subspace and mixed states supported on it.

\subsection{Parametrization and optimization}

Owing to the challenges in characterizing $\sigma_r$ (states with at most tensor rank $r$) and the non-convex nature of optimization, previous work sidestepped a direct computation of $E_r$. Instead, it focused on exploring the \textit{lower bound} of $E_r$ through mathematical derivation, or by relaxing it in a manner that allows for solution via the semi-definite programming (SDP) method, such as the positive partial transpose (PPT) relaxation:
\begin{equation}
\label{PPT}
\begin{aligned}
 E_2(\mathcal{S})&=\min _{|\psi_{\text{prod}}\rangle}\left\langle\psi_{\text{prod}}\left|\mathcal{P}_{\mathcal{S}}^{\perp}\right| \psi_{\text{prod}}\right\rangle \\
&=\min _{\left|\psi_{\text {prod}}\right\rangle} \operatorname{tr}\left[\mathcal{P}_{\mathcal{S}}^{\perp}\left|\psi_{\text {prod}}\right\rangle\left\langle\psi_{\text {prod}}\right|\right] \\
& \geqslant \min _{\substack{\rho \geqslant 0 \\
\forall_i \rho^{T_i} \geqslant 0}} \operatorname{tr}\left[\mathcal{P}_{\mathcal{S}}^{\perp} \rho\right],
\end{aligned}
\end{equation}
where $|\psi_{\text{prod}}\rangle$ denotes the \textit{fully} product state in $\sigma_1$ and $\rho^{T_i}$ denotes the partial transpose of $\rho$ with respect to the $i$-th local Hilbert space $\mathcal{H}_i$.

In the following, we will introduce a parametrization strategy to directly characterize the set $\sigma_r$ in the manifold optimization framework \cite{hu2020brief, absil2008optimization}. First, we can consider the set of \textit{unnormalized} states with tensor rank $r$, denoted by $\tilde{\sigma}_r$, for every $|\tilde{\varphi}_r\rangle \in \tilde{\sigma}_r$, it can be written as
\[
    | \tilde{\varphi}_r\rangle = \sum_{i=1}^r \tilde{\lambda_i} \vert \phi_i^{(1)}\rangle \otimes \vert \phi_i^{(2)}\rangle \otimes \cdots \otimes \vert \phi_i^{(n)}\rangle.
\]
For simplicity, we denote $\tilde{\lambda}=(\tilde{\lambda_1}, \tilde{\lambda_2}, \cdots, \tilde{\lambda_r}) \in \mathbb{R}_{+}^{r}$ and $| \phi_i \rangle= \otimes_{k=1}^{n}| \phi_i^{(k)}\rangle \in \mathcal{B}=\mathbb{S}^{2d_1-1} \times \mathbb{S}^{2d_2-1} \times \cdots \times \mathbb{S}^{2d_n-1}$, where $\mathbb{R}_{+}^{r}$ represents the $r$-dimensional manifold of positive numbers and $\mathcal{B}$ denotes the product manifold composed of a series of $(2d_k-1)$-dimensional sphere manifold $\mathbb{S}^{2d_k-1}$ ($d_k$ is the local dimension of $k$th Hilbert space). Then the original problem (Eq.~(\ref{gm-r-subspace})) can be formulated numerically as an optimization over the manifold $\mathcal{M}_{r-1}$

\[
    E_r(\mathcal{S})= \min _{x \in \mathcal{M}_{r-1}}\left\langle\varphi_{r-1}(x)\left|\mathcal{P}_{\mathcal{S}}^{\perp}\right| \varphi_{r-1}(x)\right\rangle= \min_{x \in \mathcal{M}_{r-1}} f_{\mathcal{S}}(x).
\]
Here, $| \varphi_{r-1}(x)\rangle$ is the \textit{normalized} state from $|\tilde{\varphi}_{r-1}\rangle$ via adjusting $\tilde{\lambda}$, where $x=(\tilde{\lambda}, \phi_1, \phi_2, \cdots, \phi_{r-1}) \in \mathcal{M}_{r-1}= \mathbb{R}^{r-1}_{+} \times \mathcal{B}^{r-1}$. And here, $f_{\mathcal{S}}$ is a simplified notation for the related function determined by $\mathcal{S}$.

\textit{Trivialization} \cite{lezcanocasado2019trivializations} provides a fresh and efficient perspective for manifold optimization, which requires a mapping $g$ from the free Euclidean space to the manifold $\mathcal{M}_{r-1}$, such that
\[
    E_r(\mathcal{S})=\min_{x \in \mathcal{M}_{r-1}} f_{\mathcal{S}}(x)=\min_{\vec{\theta} \in \mathbb{R}^{D}}f_{\mathcal{S}}[g(\vec{\theta})].
\]
The problem is now transferred into an unconstrained optimization problem over the $D$-dimensional Euclidean space.
Thus, the remaining task is to find such a satisfying mapping $g$. Since the manifold $\mathcal{M}_r$ is a product manifold composed of $\mathbb{R}^{r}_{+}$ and $\mathcal{B}^r$, we can consider them separately. For the positive numbers $\tilde{\lambda} \in \mathbb{R}_{+}^r$, we can use the SoftPlus mapping \cite{softplus} to ensure the positivity
\[
    \tilde{\lambda}=\log(1+e^{\vec{\gamma}}),
\]
where $\vec{\gamma}$ is a $r$-dimensional real vector. As for each normalized states $|\phi_i^{(k)}\rangle \in \mathbb{S}^{2d_k-1}$, we can obtain them by mapping some $d_k$-dimensional real vectors $\vec{\alpha_i}, \vec{\beta_i}$ onto the sphere manifold $\mathbb{S}^{2d_k-1}$ as follows
\[ |\phi_{i}^{(k)}\rangle=\frac{\vec{\alpha_i}+\mathrm{i}\vec{\beta_{i}}}{\left\Vert \vec{\alpha_i}+\mathrm{i}\vec{\beta_{i}}\right\Vert _{2}}. \]
Given these two fundamental mappings, finally we are able to map a free Euclidean space $\mathbb{R}^D$ onto the manifold $\mathcal{M}_{r-1}$ with the dimension
\[
    D = (2\sum_k d_k+1)(r-1).
\]

As for the projector $\mathcal{P}_{\mathcal{S}}^{\bot}$, it can be constructed as
\[ \mathcal{P}_{\mathcal{S}}^{\bot}=I- \mathcal{P}_{\mathcal{S}}= I-\sum_{i=1}^{d_{\mathcal{S}}}|e_{i}\rangle\langle e_{i}|, \]
where $\{|e_i\rangle\}$ is the orthonormal basis of the given subspace $\mathcal{S}$ that can be obtained through Gram-Schmidt process. Then we can define a loss function $\mathcal{L}_r(\vec{\theta}; \mathcal{S})=\langle \varphi_r(\vec{\theta})|\mathcal{P}_{\mathcal{S}}^{\bot}| \varphi_r(\vec{\theta})\rangle$ to compute $E_r(\mathcal{S})$. The relevant algorithm is summarized in Algorithm~\ref{GM}.

\RestyleAlgo{ruled}
\begin{algorithm}[H]
\caption{Geometric measure of $r$-bounded rank $E_r$}
\label{GM}
\KwIn{A subspace $\mathcal{S} \subset \bigotimes_{k=1}^{n}\mathcal{H}_k$ with $\text{dim}(\mathcal{H}_k)=d_k$, spanned by $\{\vert \psi_1\rangle, \vert \psi_2\rangle, \dots, \vert \psi_m\rangle\}$, the level of entanglement $r$, converge tolerance $\epsilon$, the number of trials $N$}
\KwOut{Geometric measure of $r$-bounded rank $E_r(\mathcal{S})$}

Orthogonalize the set $\{\vert \psi_1\rangle, \vert \psi_2\rangle, \dots, \vert \psi_m\rangle\}$ through the Gram-Schmidt process, then obtain the projector of the subspace $\mathcal{P}_{\mathcal{S}}$

$\mathcal{P}_{\mathcal{S}}^{\bot}=I-\mathcal{P}_{\mathcal{S}}$

$t \leftarrow 1$

\While{$t \leq N$}{Initialize a random real vector $\vec{\theta} \in \mathbb{R}^{D}$, $D=(2\sum_k d_k+1)(r-1)$

\While{$E_r$ has not converged under tolerance $\epsilon$}{
    Compute $\mathcal{L}_{r-1}\left(\vec{\theta};\mathcal{S}\right)=\langle\varphi_{r-1}(\vec{\theta})|\mathcal{P}_{\mathcal{S}}^{\bot}|\varphi_{r-1}(\vec{\theta})\rangle$

    Update $\vec{\theta}$ based on the gradient descent
}
$E^{(t)}_r=\mathcal{L}_{r-1}\left(\vec{\theta};\mathcal{S}\right)$

$t \leftarrow t+1$
}
$E_r(\mathcal{S})=\min\{E_r^{(1)},\dots,E_r^{(N)}\}$

\end{algorithm}

To minimize the loss function, we adopt the gradient-based optimization method with the gradients obtained from gradient back-propagation provided in Pytorch deep learning framework \cite{pytorch}. Different from the common neural networks, there is no randomness in our loss function, so the classical optimization method, L-BFGS-B implemented in SciPy package \cite{virtanen2020scipy}, should converge faster than those commonly used in training neural networks (e.g. SGD, Adam). However, gradient-based methods do not guarantee convergence to the global minimum, which provide \textit{upper bounds} in principle. In order to obtain accurate values, we employ a probabilistic approach: minimizing the loss with randomly initialized parameters across $N$ trials. Here, $N$ can be considered as a hyper-parameter in the optimization process.

\section{Results}\label{results}
In this section, we demonstrate the application of the proposed method across various scenarios, including both bipartite and multipartite cases. We primarily compare our results with other established methods, such as the PPT relaxation \cite{demianowicz2019entanglement,zhang2020numerical} described in Eq.~(\ref{PPT}) and the hierarchical method \cite{johnston2022complete}. Additionally, we compare the seesaw strategy \cite{streltsov2011simple} with our approach, which also provides upper bounds for GME. For each optimization, we fix the trial number $N=3$ and the tolerance $\epsilon=10^{-10}$ in the optimization. Numerical results show that this choice is effective enough but may not be optimal for certain cases.
\subsection{Bipartite cases}

\subsubsection{Subspace with analytically known $E_r$}
\textit{Example 1}. The subspace $\mathcal{S}_{2\times d}^{\theta} \subset \mathcal{H}_A \otimes \mathcal{H}_B$ with $d_A=2$ and $d_B=d$, is given by the span of the following vectors \cite{demianowicz2019entanglement}:
\begin{equation}
    \left|\psi_i\right\rangle_{A B}=a|0\rangle_A\left|i\right\rangle_B+b|1\rangle_A\left|i+1\right\rangle_B,
    \label{eq:analytic_subspace}
\end{equation}
where $i=0,1,\cdots,d-2$, with $a=\cos(\theta/2)$ and $b=\exp(i\xi)\sin(\theta/2)$, $\theta \in (0,\pi)$, $\xi \in [0,2\pi)$.

Clearly, the dimension of $\mathcal{S}_{2\times d}^{\theta}$ is $d-1$, which is also the maximal available dimension of an entangled subspace in this scenario. Furthermore, it has been proved that
\[
E_{2}\left(\mathcal{S}_{2 \times d}^\theta\right)=\frac{1}{2}\left[1-\sqrt{1-\sin ^2 \theta \sin ^2\left(\frac{\pi}{d}\right)}\right].
\]

\begin{figure}[H]
    \centering
    \includegraphics[width=\linewidth]{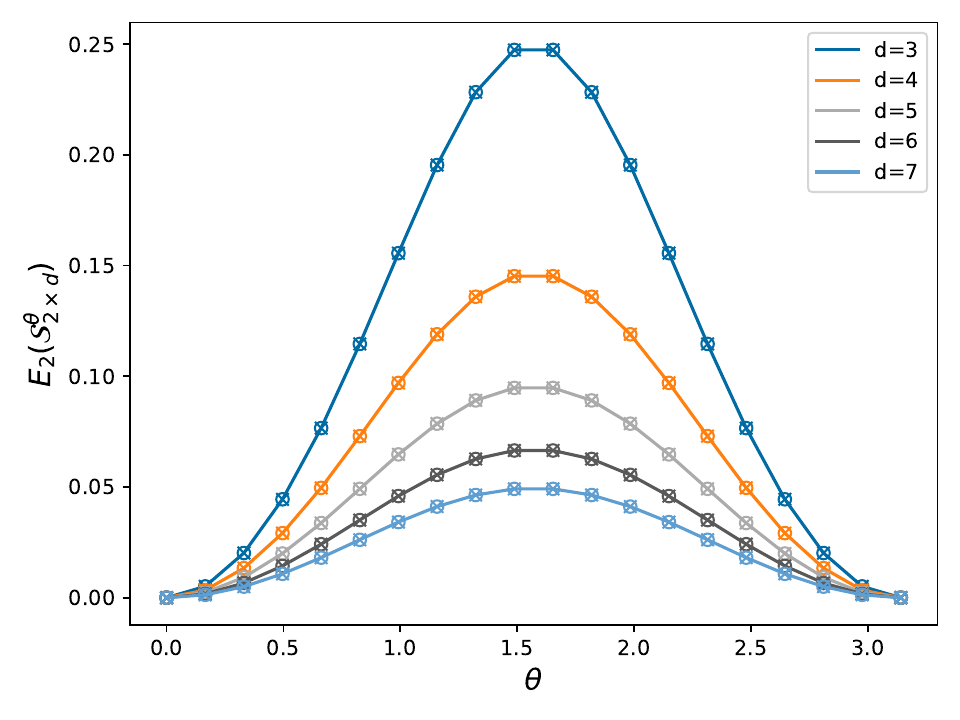}
    \caption{$E_2(\mathcal{S}_{2\times d}^{\theta})$ as a function of $\theta$. Different lines denote the analytical results of different dimensions, circles denote the results obtained by gradient descent, and crosses represent the results from the PPT relaxation. Numerical results from both two different methods match the analytical results.}
    \label{fig:analytic_gm}
\end{figure}

In Fig.~\ref{fig:analytic_gm}, we plot $E_2(\mathcal{S}_{2\times d}^{\theta})$ as a function of $\theta$ for different dimensions and compare the analytical results with those obtained via the gradient descent and PPT relaxation methods. For convenience, we fix $\xi=0$. As observed, for this kind of low-dimensional entanglement, both methods provide accurate results.

\begin{figure}[H]
    \centering
    \includegraphics[width=\linewidth]{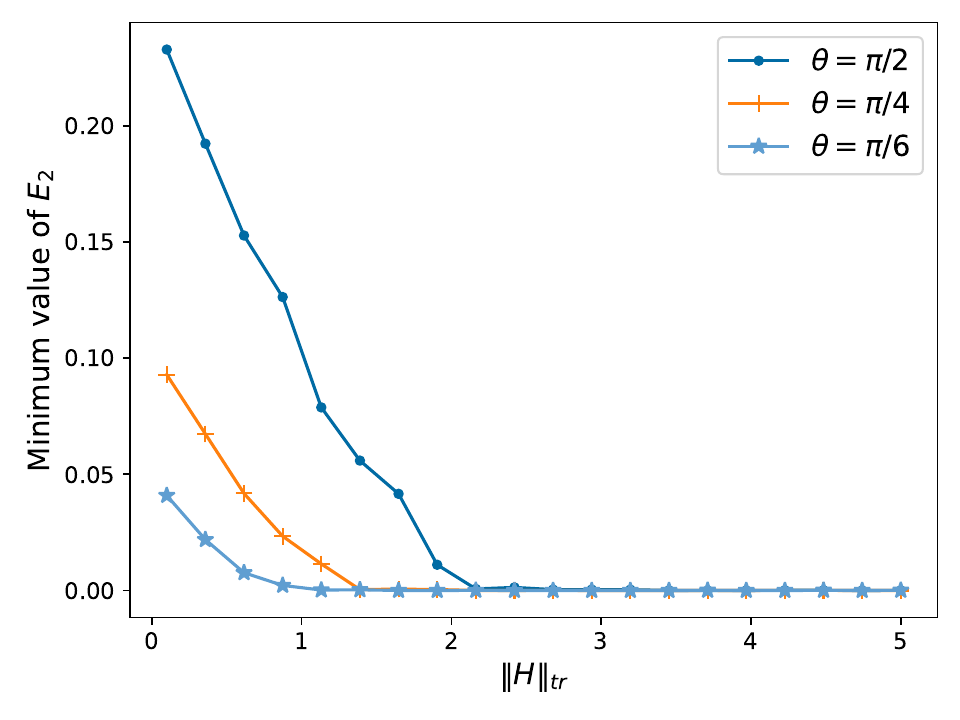}
    \caption{Minimum values of $E_2$ after the random perturbations $U=e^{-iH}$ versus different trace norm of $H$. A nonzero value of $E_2$ means the perturbated subspace maintains its entanglement. Larger $E_2(\mathcal{S})$ brings more robustness as expected.}
    \label{fig:robustness}
\end{figure}

The robustness of different entangled subspaces can also be explored, as depicted in Fig.~\ref{fig:robustness}. Three different subspaces ($\theta=\frac{\pi}{2}, \frac{\pi}{4}, \frac{\pi}{6})$ with $d=3$ are selected for comparison. We randomly generate 1000 unitary perturbations $U=e^{-iH}$ for each different trace norm of $H$. Utilizing gradient descent strategy, we compute the minimum geometric measure $E_2$ after the perturbations with different trace norms. A non-zero value of $E_2$ indicates that the subspace remains entangled after perturbations. As Theorem 1 proves, entangled subspaces with larger $E_2$ values are more robust against perturbations.

\subsubsection{Bound and high-dimensional entanglement}
Although the PPT relaxation can give accurate $E_2$ for most subspaces in practice, it fails when faced with bound entanglement and cannot estimate the geometric measure of higher bounded rank, i.e., $E_r(r>2)$.

\textit{Example 2}. One famous approach to construct bound entangled states is using the unextendible product basis (UPB) \cite{bennett1999unextendible}, as follows
\[
\rho = \frac{1}{d_Ad_B-d_{\mathcal{S}}}\mathcal{P}_{\mathcal{S}}^{\bot},
\]
where $d_A, d_B$ are the local dimensions for bipartite states, $\mathcal{S}$ represents the subspace spanned by the given UPB and $d_{\mathcal{S}}$ is the dimension of that subspace. For example, we can consider the five-state tiles UPB \cite{divincenzo2003unextendible} (here we omit normalization for brevity),
\[
\begin{aligned}
\mathcal{S}_{\text {tiles }}= & \operatorname{span}\{|0\rangle \otimes(|0\rangle-|1\rangle),|2\rangle \otimes(|1\rangle-|2\rangle),\\ &(|0\rangle-|1\rangle)
 \otimes|2\rangle,(|1\rangle-|2\rangle) \otimes|0\rangle,(|0\rangle+|1\rangle+|2\rangle) \\
& \otimes(|0\rangle+|1\rangle+|2\rangle)\} \subset \mathcal{H}_A \otimes \mathcal{H}_B .
\end{aligned}
\]

As we have mentioned before, the geometric measure of the support space can give a lower bound for the geometric measure of the state, i.e., $E_r(\rho) \geq E_r[\text{supp}(\rho)]$. For the state $\rho_{\text{tiles}}$ constructed by the tiles UPB, we can estimate $E_2[\text{supp}(\rho_{\text{tiles}})]$ by the PPT relaxation, which gives around $10^{-12}$ close to $0$ while the gradient descent gives around $0.0284$. That means  PPT cannot detect bound entanglement while the gradient descent does.

\textit{Example 3}. Let $d_A=d_B=4$ and consider the following mixed state from \cite{johnston2022complete}:
\[
\rho=\frac{1}{3} \sum_{i=1}^3\left|\psi_i\right\rangle\left\langle \psi_i\right| \in \mathcal{H}_A \otimes \mathcal{H}_B,
\]
where (we omit normalization for brevity)
\[
\begin{aligned}
& \left|\psi_1\right\rangle=|0\rangle \otimes|0\rangle+|1\rangle \otimes|1\rangle+|2\rangle \otimes|2\rangle+|3\rangle \otimes|3\rangle, \\
& \left|\psi_2\right\rangle=|0\rangle \otimes|1\rangle+|1\rangle \otimes|2\rangle+|2\rangle \otimes|3\rangle+|3\rangle \otimes|0\rangle, \\
& \left|\psi_3\right\rangle=|0\rangle \otimes|2\rangle+|1\rangle \otimes|3\rangle+|2\rangle \otimes|0\rangle-|3\rangle \otimes|1\rangle .
\end{aligned}
\]

By using the PPT relaxation, we figure out that $E_2(\rho) \geq 0.382367$, indicates the presence of entanglement. Further, by computing the geometric measure of higher bounded rank via gradient descent, we can conclude that $\rho$ is high-dimensional entangled since $E_3(\rho) \geq 0.06558$, i.e., it can only be an ensemble contains at least rank-3 entangled states. This information is not available for the PPT relaxation.

\subsection{Multipartite cases}
\subsubsection{Border rank of pure states}
The hierarchical method can also detect bound or high-dimensional entanglement in the \textit{bipartite} scenario. However, both the PPT relaxation and hierarchical method fall short when certifying high-dimensional entanglement in the \textit{multipartite} settings even for a single multipartite pure state $|\psi\rangle$. Conversely, through gradient descent, we can efficiently obtain the value of $E_r(|\psi\rangle)$, which provides information about the high-dimensional entanglement, i.e., border rank of the given state.

\textit{Example 4}. Permutation symmetric states in $n$-qubit system are given by the Dicke states, which are mathematically expressed as the sum of all permutations of computational basis states with $n-k$ qubits being $|0\rangle$ and $k$ being $|1\rangle$:
\[
\left|D_n^k\right\rangle=\left(\begin{array}{l}
n \\
k
\end{array}\right)^{-1 / 2} \sum_{\text {perm }} \underbrace{|0\rangle|0\rangle \cdots|0\rangle}_{n-k} \underbrace{|1\rangle|1\rangle \cdots|1\rangle}_k,
\]
with $0 \leq k \leq n$. It is proved that the closest product state in $\sigma_1$ for computing $E_2$ \cite{wei2003geometric, wei2010matrix} has the form
\begin{equation}
\label{eq: closest_state}
    |\varphi\rangle=\left(\sqrt{\frac{n-k}{n}}|0\rangle+\sqrt{\frac{k}{n}}|1\rangle\right)^{\otimes n},
\end{equation}
i.e. a tensor product of $n$ identical single qubit states. From that, the geometric measure is found to be
\[
E_2(|D_n^k\rangle)=1-\binom{n}{k}\left(\frac{k}{n}\right)^k\left(\frac{n-k}{n}\right)^{n-k}.
\]

Due to the symmetry of Dicke states, we can just discuss the cases where $k \leq \left\lfloor\frac{n}{2}\right\rfloor$. It has been proved that for a Dicke state $|D_n^k\rangle$, its border rank $\underline{R}=k+1$ \cite{gharahi2021algebraic}, which means there should be a transition between zero and nonzero for $E_{k+2}$ and $E_{k+1}$.

\begin{figure}[H]
    \centering
    \includegraphics[width=\linewidth]{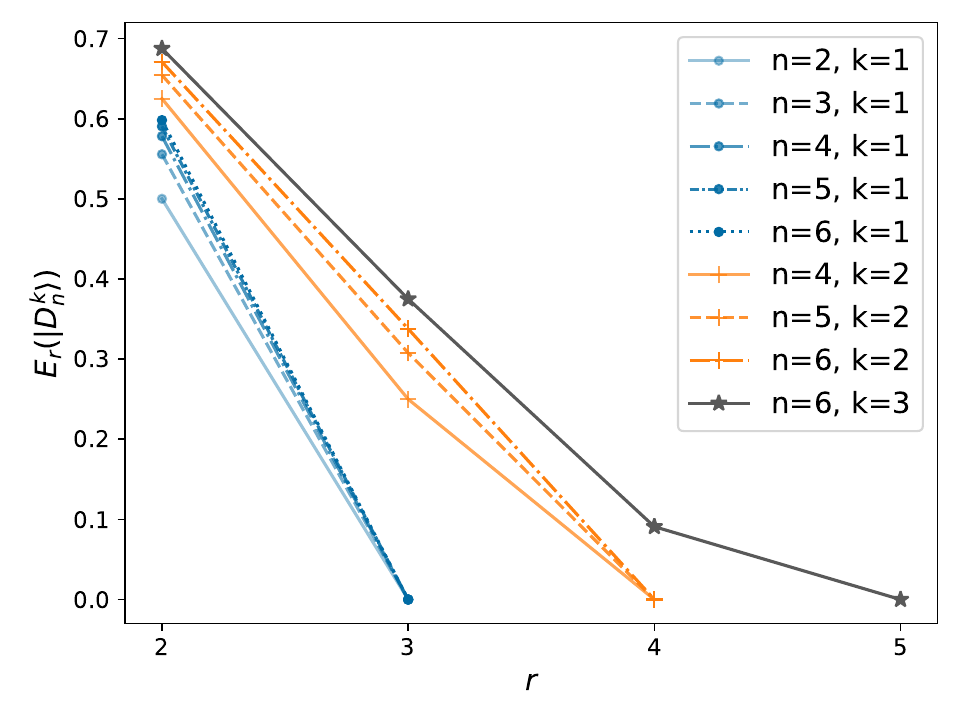}
    \caption{$E_r(|D_n^k\rangle)$ as a function of $r$ for different $n$ and $k$. The transition of $E_r$ between zero and nonzero at $k+1$ and $k+2$ implies the border rank of $|D_n^k\rangle$ is $k+1$ as predicted.}
    \label{fig:border_rank}
\end{figure}

We compute the geometric measure $E_r(|D_n^k\rangle)$ for different $n$ and $k$, as shown in Fig.~\ref{fig:border_rank}. For $r=2$, we compare the numerical results with analytical results, which implies the error is close to the machine precision. The analytical form in Eq.~(\ref{eq: closest_state}) can also be found in the optimization process. Generally, uncovering the closest state proves challenging unless the given state exhibits certain symmetries \cite{hubener2009geometric}. Our methodology furnishes a potent instrument for the exploration of these closest states even in $\sigma_r (r>2)$. Moreover, we observe distinct transitions from non-zero to zero for different Dicke states, indicating the border rank of the given $|D_n^k\rangle$ as postulated. In Appendix~\ref{see-saw}, we compare the results obtained using the seesaw strategy and the PPT relaxation for calculating \( E_2(|D_n^k\rangle) \). We find that the time efficiency of gradient descent lies between these two methods. However, as we observe, neither the seesaw strategy nor the PPT relaxation can detect higher-dimensional entanglement, which highlights the major advantage of the gradient descent approach we propose.

Border rank and tensor rank are also closely related to \textit{algebraic complexity theory}. One of the most important open questions in computer science is to understand the computational complexity of \textit{matrix multiplication}. One would like to know how many multiplication operations are required in order to multiply $n \times n$ matrices. The naive approach uses $n^3$ multiplications while Strassen found an algorithm only needs seven multiplications for $2\times2$ matrices \cite{strassen1969gaussian}.

The precise number of least multiplications for the matrix multiplication is called the \textit{rank} of matrix multiplication while the concept of \textit{border rank} arises when some matrix multiplications can be approximated with arbitrary precision by less complicated multiplications, which have lower ranks \cite{bini1979n2,bini1980approximate}. These approximations can give faster algorithms for matrix multiplication in practice.

\textit{Example 5}. It turns out that the rank of $n \times n$ matrix multiplication is equivalent to the rank of the following tripartite state (unnormalized) \cite{chitambar2008tripartite,christandl2023resource}:
\[
\begin{aligned}
    \left|\Phi_n\right\rangle_{ABC}&=|\Phi\rangle_{A B}|\Phi\rangle_{A C}|\Phi\rangle_{B C}\\
&=\sum_{i, j, k=0}^{n-1}|i j\rangle_A|i k\rangle_B|j k\rangle_C,
\end{aligned}
\]
where any two parties share an EPR pair $|\Phi\rangle$ of dimension $n$. By computing the values of $E_r(\left|\Phi_2\right\rangle_{ABC})$ for different $r$, we find that $E_7$ is very close to $\frac{1}{8}$ while $E_8$ is around $10^{-14}$, which implies that $\underline{R}(\left|\Phi_2\right\rangle_{ABC})=7$. This numerical result matches the previous finding \cite{landsberg2006border}, validating the practicability of our approach for studying border rank in many relevant fields, particularly in mathematics and computer science.

\subsubsection{Completely entangled subspaces}
\textit{Example 6}. Consider the following four three-qubit states \cite{branciard2010evaluation}:
\[
\begin{aligned}
\left|\psi_0\right\rangle&=|0\rangle_A|0\rangle_B|0\rangle_C, \\
\left|\psi_1\right\rangle&=|1\rangle_A|+\rangle_B|-\rangle_C, \\
\left|\psi_2\right\rangle&=|-\rangle_A|1\rangle_B|+\rangle_C, \\
\left|\psi_3\right\rangle&=|+\rangle_A|-\rangle_B|1\rangle_C.
\end{aligned}
\]
They form a UPB, which means no product state can be found in the orthogonal space of the subspace $\mathcal{S}$ spanned by them. In other words, $\mathcal{S}^{\bot}$ is a completely entangled subspace. However, similar to the bipartite UPB case, PPT relaxation cannot detect its entanglement since the lower bound of $E_2$ obtained close to $10^{-14}$ while the gradient descent gives the value around $0.08144$, close to the analytical value $1-\frac{3\sqrt{6}}{8}$.

\textit{Example 7}. The largest possible dimension of a completely entangled subspace (CES) of $\mathcal{H}_1 \otimes \mathcal{H}_2 \otimes \mathcal{H}_3$ is $d_1 d_2 d_3-d_1-d_2-d_3+2$ and one particular example of such a subspace is \cite{bhat2006completely}
\[
\begin{aligned}
    \mathcal{S}=\operatorname{span}\{ &|i_1\rangle \otimes |i_2\rangle \otimes |i_3\rangle-|j_1\rangle \otimes |j_2\rangle \otimes |j_3\rangle: \\
    &\sum_{r=1}^3 i_r=\sum_{r=1}^3 j_r, 0 \leq i_r \leq d_r-1, 1 \leq r \leq 3 \}.
\end{aligned}
\]
Our method is able to certify these maximal-dimension CES for different values of $d_1, d_2$ and $d_3$ quite efficiently. We calculate the computational time for certifying these CESs via different methods in Table. \ref{table: ces}.

\begin{table}[H]
    \centering
    \caption{\label{table: ces} Computational time for certifying CESs via different methods, including Hier (hierarchical method), PPT (PPT relaxation), and GD (gradient descent). The values of geometric measure $E_2$ obtained by PPT and GD are also presented. ``-'' means it is not available in acceptable time.}
    \begin{tabular}{c|ccc|cc}
        \toprule
        \multirow{2}*{$(d_1,d_2,d_3)$} & \multicolumn{3}{c|}{Time(s)} & \multicolumn{2}{c}{Geometric measure} \\
        \cmidrule{2-4} \cmidrule{5-6}
          & Hier  & PPT  & GD & $E_2^{\text{PPT}}$ & $E_2^{\text{GD}}$\\
        \midrule
        (2,2,2) & 0.01   & 0.09     & 0.07  & $2.00\times 10^{-1}$    & $2.50\times 10^{-1}$ \\
        (2,2,4) & 0.07 & 0.16 & 0.08 & $4.49\times 10^{-2}$    & $4.50\times 10^{-2}$ \\
        (2,2,6) & 0.30 & 0.33 & 0.10 & $1.23\times 10^{-2}$    & $1.23\times 10^{-2}$ \\
        (2,3,4) & 180.42 & 0.32 & 0.08 & $1.41\times 10^{-2}$    & $1.41\times 10^{-2}$ \\
        (2,3,6) & - & 1.07 & 0.15 & $2.86\times 10^{-3}$    & $2.86\times 10^{-3}$ \\
        (2,3,8) & - & 2.10 & 0.21 & $7.62\times 10^{-4}$    & $7.62\times 10^{-4}$ \\
        (3,3,6) & - & 2.71 & 0.18 & $7.20\times 10^{-4}$    & $7.20\times 10^{-4}$ \\
        (3,3,8) & - & 8.55 & 0.32 & $1.57\times 10^{-4}$    & $1.57\times 10^{-4}$ \\
        (3,4,7) & - & 18.27 & 0.40 & $7.98\times 10^{-5}$    & $7.98\times 10^{-5}$ \\
        (4,4,7) & - & 189.31 & 0.55 & $2.02\times10^{-5}$ & $2.02\times10^{-5}$ \\
        (4,5,10) & - & - & 2.92 & - & $3.54\times10^{-7}$ \\

        \bottomrule
    \end{tabular}
\end{table}

As we can see, the hierarchical method is most easily limited by the dimension sizes. Although the PPT method can detect most of them successfully and give the accurate lower bounds of $E_2$ for relatively small dimensions, its computational time will increase significantly since it requires a quantum state $\rho$ with the dimension $d_A d_B d_C \times d_A d_B d_C$. On the contrary, the gradient descent approach can deal with much larger dimensions due to the need for only $2(d_A+d_B+d_C)+1$ real parameters. This example shows the practicability and efficiency of our approach clearly.

\subsubsection{Genuinely entangled subspaces}
\textit{Example 8}. The entangled subspace spanned by the vectors in Eq.~(\ref{eq:analytic_subspace}) can be generalized to genuinely entangled subspace (GES) $\mathcal{S}_{2 \times d^{2}}^\theta$ in $\mathcal{H}_A \otimes\mathcal{H}_B\otimes \mathcal{H}_C$ , spanned by
\[
|\psi_{i_1 i_2} \rangle_{ABC}=a|0\rangle_A|i_1\rangle_B |i_2\rangle_C + b|1\rangle_A|i_1+1\rangle_B |i_2+1\rangle_C,
\]
where $d_A=2, d_B=d_C=d$, $i_1,i_2 = 0,1,\dots,d-2$ and $a=\cos(\theta/2), b=e^{\xi}\sin(\theta/2), \theta \in (0,\pi), \xi \in [0, 2\pi)$.

For this particular genuinely entangled subspace, it has been proved that \cite{demianowicz2019entanglement}, for any possible bipartition $K|K^c$ in $\mathcal{H}_A \otimes\mathcal{H}_B\otimes \mathcal{H}_C$:
\[
E^{K|K^c}_{2}\left(\mathcal{S}_{2 \times d^{2}}^\theta\right)=\frac{1}{2}\left[1-\sqrt{1-\sin ^2 \theta \sin ^2\left(\frac{\pi}{d}\right)}\right].
\]
We compare this analytical result with ones obtained by gradient descent, which implies the errors are close to the machine precision. That means our approach can successfully detect their genuine entanglement and give accurate values for geometric measure over bipartitions.

\section{Conclusions and Outlooks}\label{conclusions}

In this work, we introduce a potent tool for examining the entanglement within a given subspace. Through several examples, the universality, practicability, and efficiency of our approach become evident. The core idea of our methodology involves finding a suitable way to map a free Euclidean space onto the complicated manifold in the given quantum scenario. Rather than augmenting the dimension of quantum problems to supplement convexity, we employ the highly developed non-convex optimization techniques from machine learning to expedite the process. Consequently, we can address various issues related to quantum subspace, ranging from the certification of bipartite entangled subspaces, to the computation of the border rank of multipartite states, and the verification of completely and genuinely entangled subspaces. All of these results can be reproduced in our public repository \cite{zhang_2024_12166896}.

Convex optimization techniques, particularly semi-definite programming (SDP) method, are often used for quantifying entanglement. Although SDP can guarantee a global minimum, mathematically, it still only provides lower bounds. This means that when it returns zero, we cannot exclude the presence of entanglement, such as bound entanglement. Moreover, as highlighted by the results compared in the paper, we can see that its computational complexity increases significantly with dimension, implying that it may only provide entanglement information for relatively small quantum systems.

In contrast, algorithms based on gradient descent strategies provide upper bounds. These also have their drawbacks and advantages. For example, when they return a non-zero value, we theoretically cannot exclude the possibility of a local minimum. However, as we have observed, they offer a more universal and versatile approach, enabling application to a broader range of scenarios, such as the detection of high-dimensional entanglement. Furthermore, thanks to the rapid advancement of non-convex optimization techniques in recent years, especially those applied in machine learning, their computational efficiency significantly surpasses that of SDP, making it feasible to address quantum problems in more complex systems.

In practice, these two approaches do not conflict. To calculate actual values, we always need both upper and lower bounds to better estimate them. We hope that our research can inspire more applications of non-convex optimization in quantum information.

\acknowledgments
We gratefully acknowledge Ningping Cao, Shu Zhou, Jiahao Hu, Kuigang She and Yiu Tung Poon for insightful discussions and the assistance of ChatGPT in facilitating the writing process. X-R. Zhu, C. Zhang and B. Zeng are supported by GRF grant No. 16305121.

\appendix
\section{Comparison of time and accuracy for computing the geometric measure of entanglement for Dicke states}
\label{see-saw}

Here, we introduce the seesaw strategy proposed in Ref.~\cite{streltsov2011simple} for the calculation of geometric measure of entanglement $E_G$ [see Eq.~(\ref{GME})].

Given a pure state $|\psi\rangle$ in an $n$-partite Hilbert space $\bigotimes_{i=1}^n \mathcal{H}_i$, one straightforward strategy to find the closest fully product state $|\varphi\rangle$ under the geometric measure is as follows: 

Start with a random fully product state $|\varphi_0\rangle = \bigotimes_{i=1}^n |\phi_0^{(i)}\rangle$. Then, consider the unnormalized state $|\tilde{\psi}\rangle = \left(\bigotimes_{i=2}^n \langle \phi_0^{(i)}|\right)|\psi\rangle$. To minimize $E_G$, i.e., to maximize the overlap $|\langle \varphi|\psi\rangle|$ for fixed states $\bigotimes_{i=2}^n |\phi_0^{(i)}\rangle$, replace $|\phi_0^{(1)}\rangle$ with the state $|\phi_1^{(1)}\rangle = \frac{|\tilde{\psi}\rangle}{\sqrt{\langle \tilde{\psi}|\tilde{\psi}\rangle}}$.
\vspace{0.1cm}
\begin{table}[H]
    \centering
    \begin{tabular}{c|ccccccc}
        \toprule
        \multirow{2}{*}{$(n,k)$} & \multirow{2}{*}{$E_G$} & \multicolumn{2}{c}{GD} & \multicolumn{2}{c}{seesaw} &PPT\\
        & & error & time & error & time & time \\
        \midrule
        (5,1) & 0.5904 & $0.21$ & 0.0101 & $0.23$ & 0.103 & 1.73 \\
        (5,2) & 0.6544 & $0.03$ & 0.087 & $0.78$ & 0.0091 & 1.64\\
        (5,3) & 0.6544 & $0.13$ & 0.084 & $0.70$ & 0.0091 & 1.59\\
        (6,1) & 0.5981 & $0.21$ & 0.129 & $0.27$ & 0.017 & 25.1\\
        (6,2) & 0.6708 & $0.02$ & 0.122 & $0.44$ & 0.018 & 26.5\\
        (6,3) & 0.6875 & $0.12$ & 0.166 & $0.65$ & 0.018 & 25.7\\
        (7,1) & 0.6034 & $0.01$ & 0.173 & $1.08$ & 0.019 & NA\\
        (7,2) & 0.6813 & $0.22$ & 0.163 & $0.24$ & 0.019 & NA\\
        (7,3) & 0.7062 & $0.04$ & 0.150 & $0.16$ & 0.020 & NA\\
        \bottomrule
    \end{tabular}
    \caption{\label{table: dicke}Comparison of gradient descent (GD), seesaw strategy and the PPT relaxation method in calculating the geometric measure of entanglement for various Dicke states $|D_n^k\rangle$. Time is measured in seconds. The error and time columns denote the absolute error (in unit $\times 10^{-12}$) compared to the analytical value $E_G$ and computational time for the corresponding method. NA means not available within acceptable time.}
\end{table}

This process is then repeated for each of the other parties. This iterative method, known as the seesaw strategy, is continued until convergence. The convergence criterion is set such that the fidelity of every local state before and after the update is greater than $1 - 10^{-10}$. This strategy can also provide an upper bound for the geometric measure of entanglement.

In Table.~\ref{table: dicke}, we compare the computational time and accuracy for calculating the geometric measure of entanglement for Dicke states $|D_n^k\rangle$ using the seesaw strategy, the PPT relaxation, and gradient descent (GD). Here, for each state, we repeat the optimization three times for both seesaw and gradient descent strategy.

\bibliography{reference.bib}

\begin{thebibliography}{74}%
\makeatletter
\providecommand \@ifxundefined [1]{%
 \@ifx{#1\undefined}
}%
\providecommand \@ifnum [1]{%
 \ifnum #1\expandafter \@firstoftwo
 \else \expandafter \@secondoftwo
 \fi
}%
\providecommand \@ifx [1]{%
 \ifx #1\expandafter \@firstoftwo
 \else \expandafter \@secondoftwo
 \fi
}%
\providecommand \natexlab [1]{#1}%
\providecommand \enquote  [1]{``#1''}%
\providecommand \bibnamefont  [1]{#1}%
\providecommand \bibfnamefont [1]{#1}%
\providecommand \citenamefont [1]{#1}%
\providecommand \href@noop [0]{\@secondoftwo}%
\providecommand \href [0]{\begingroup \@sanitize@url \@href}%
\providecommand \@href[1]{\@@startlink{#1}\@@href}%
\providecommand \@@href[1]{\endgroup#1\@@endlink}%
\providecommand \@sanitize@url [0]{\catcode `\\12\catcode `\$12\catcode
  `\&12\catcode `\#12\catcode `\^12\catcode `\_12\catcode `\%12\relax}%
\providecommand \@@startlink[1]{}%
\providecommand \@@endlink[0]{}%
\providecommand \url  [0]{\begingroup\@sanitize@url \@url }%
\providecommand \@url [1]{\endgroup\@href {#1}{\urlprefix }}%
\providecommand \urlprefix  [0]{URL }%
\providecommand \Eprint [0]{\href }%
\providecommand \doibase [0]{https://doi.org/}%
\providecommand \selectlanguage [0]{\@gobble}%
\providecommand \bibinfo  [0]{\@secondoftwo}%
\providecommand \bibfield  [0]{\@secondoftwo}%
\providecommand \translation [1]{[#1]}%
\providecommand \BibitemOpen [0]{}%
\providecommand \bibitemStop [0]{}%
\providecommand \bibitemNoStop [0]{.\EOS\space}%
\providecommand \EOS [0]{\spacefactor3000\relax}%
\providecommand \BibitemShut  [1]{\csname bibitem#1\endcsname}%
\let\auto@bib@innerbib\@empty
\bibitem [{\citenamefont {Steane}(1998)}]{steane1998quantum}%
  \BibitemOpen
  \bibfield  {author} {\bibinfo {author} {\bibfnamefont {A.}~\bibnamefont
  {Steane}},\ }\bibfield  {title} {\bibinfo {title} {Quantum computing},\
  }\href {https://doi.org/10.1088/0034-4885/61/2/002} {\bibfield  {journal}
  {\bibinfo  {journal} {Rep. Prog. Phys.}\ }\textbf {\bibinfo {volume} {61}},\
  \bibinfo {pages} {117} (\bibinfo {year} {1998})}\BibitemShut {NoStop}%
\bibitem [{\citenamefont {Preskill}(2018)}]{preskill2018quantum}%
  \BibitemOpen
  \bibfield  {author} {\bibinfo {author} {\bibfnamefont {J.}~\bibnamefont
  {Preskill}},\ }\bibfield  {title} {\bibinfo {title} {Quantum computing in the
  nisq era and beyond},\ }\href {https://doi.org/10.22331/q-2018-08-06-79}
  {\bibfield  {journal} {\bibinfo  {journal} {Quantum}\ }\textbf {\bibinfo
  {volume} {2}},\ \bibinfo {pages} {79} (\bibinfo {year} {2018})}\BibitemShut
  {NoStop}%
\bibitem [{\citenamefont {Gisin}\ \emph {et~al.}(2002)\citenamefont {Gisin},
  \citenamefont {Ribordy}, \citenamefont {Tittel},\ and\ \citenamefont
  {Zbinden}}]{gisin2002quantum}%
  \BibitemOpen
  \bibfield  {author} {\bibinfo {author} {\bibfnamefont {N.}~\bibnamefont
  {Gisin}}, \bibinfo {author} {\bibfnamefont {G.}~\bibnamefont {Ribordy}},
  \bibinfo {author} {\bibfnamefont {W.}~\bibnamefont {Tittel}},\ and\ \bibinfo
  {author} {\bibfnamefont {H.}~\bibnamefont {Zbinden}},\ }\bibfield  {title}
  {\bibinfo {title} {Quantum cryptography},\ }\href
  {https://doi.org/10.1103/revmodphys.74.145} {\bibfield  {journal} {\bibinfo
  {journal} {Rev. Mod. Phys.}\ }\textbf {\bibinfo {volume} {74}},\ \bibinfo
  {pages} {145} (\bibinfo {year} {2002})}\BibitemShut {NoStop}%
\bibitem [{\citenamefont {Pirandola}\ \emph {et~al.}(2015)\citenamefont
  {Pirandola}, \citenamefont {Eisert}, \citenamefont {Weedbrook}, \citenamefont
  {Furusawa},\ and\ \citenamefont {Braunstein}}]{pirandola2015advances}%
  \BibitemOpen
  \bibfield  {author} {\bibinfo {author} {\bibfnamefont {S.}~\bibnamefont
  {Pirandola}}, \bibinfo {author} {\bibfnamefont {J.}~\bibnamefont {Eisert}},
  \bibinfo {author} {\bibfnamefont {C.}~\bibnamefont {Weedbrook}}, \bibinfo
  {author} {\bibfnamefont {A.}~\bibnamefont {Furusawa}},\ and\ \bibinfo
  {author} {\bibfnamefont {S.~L.}\ \bibnamefont {Braunstein}},\ }\bibfield
  {title} {\bibinfo {title} {Advances in quantum teleportation},\ }\href
  {https://doi.org/10.1038/nphoton.2015.154} {\bibfield  {journal} {\bibinfo
  {journal} {Nat. Photonics}\ }\textbf {\bibinfo {volume} {9}},\ \bibinfo
  {pages} {641} (\bibinfo {year} {2015})}\BibitemShut {NoStop}%
\bibitem [{\citenamefont {Parthasarathy}(2004)}]{parthasarathy2004maximal}%
  \BibitemOpen
  \bibfield  {author} {\bibinfo {author} {\bibfnamefont {K.~R.}\ \bibnamefont
  {Parthasarathy}},\ }\bibfield  {title} {\bibinfo {title} {On the maximal
  dimension of a completely entangled subspace for finite level quantum
  systems},\ }\href {https://doi.org/10.1007/bf02829441} {\bibfield  {journal}
  {\bibinfo  {journal} {Proc. Math. Sci.}\ }\textbf {\bibinfo {volume} {114}},\
  \bibinfo {pages} {365} (\bibinfo {year} {2004})}\BibitemShut {NoStop}%
\bibitem [{\citenamefont {Walgate}\ and\ \citenamefont
  {Scott}(2008)}]{walgate2008generic}%
  \BibitemOpen
  \bibfield  {author} {\bibinfo {author} {\bibfnamefont {J.}~\bibnamefont
  {Walgate}}\ and\ \bibinfo {author} {\bibfnamefont {A.~J.}\ \bibnamefont
  {Scott}},\ }\bibfield  {title} {\bibinfo {title} {Generic local
  distinguishability and completely entangled subspaces},\ }\href
  {https://doi.org/10.1088/1751-8113/41/37/375305} {\bibfield  {journal}
  {\bibinfo  {journal} {J. Phys. A: Math. Theor.}\ }\textbf {\bibinfo {volume}
  {41}},\ \bibinfo {pages} {375305} (\bibinfo {year} {2008})}\BibitemShut
  {NoStop}%
\bibitem [{\citenamefont {Demianowicz}(2022)}]{demianowicz2022universal}%
  \BibitemOpen
  \bibfield  {author} {\bibinfo {author} {\bibfnamefont {M.}~\bibnamefont
  {Demianowicz}},\ }\bibfield  {title} {\bibinfo {title} {Universal
  construction of genuinely entangled subspaces of any size},\ }\href
  {https://doi.org/10.22331/q-2022-11-10-854} {\bibfield  {journal} {\bibinfo
  {journal} {Quantum}\ }\textbf {\bibinfo {volume} {6}},\ \bibinfo {pages}
  {854} (\bibinfo {year} {2022})}\BibitemShut {NoStop}%
\bibitem [{\citenamefont {Demianowicz}\ \emph {et~al.}(2021)\citenamefont
  {Demianowicz}, \citenamefont {Rajchel-Mieldzio{\'c}},\ and\ \citenamefont
  {Augusiak}}]{demianowicz2021simple}%
  \BibitemOpen
  \bibfield  {author} {\bibinfo {author} {\bibfnamefont {M.}~\bibnamefont
  {Demianowicz}}, \bibinfo {author} {\bibfnamefont {G.}~\bibnamefont
  {Rajchel-Mieldzio{\'c}}},\ and\ \bibinfo {author} {\bibfnamefont
  {R.}~\bibnamefont {Augusiak}},\ }\bibfield  {title} {\bibinfo {title} {Simple
  sufficient condition for subspace to be completely or genuinely entangled},\
  }\href {https://doi.org/10.1088/1367-2630/ac2a5c} {\bibfield  {journal}
  {\bibinfo  {journal} {New J. Phys.}\ }\textbf {\bibinfo {volume} {23}},\
  \bibinfo {pages} {103016} (\bibinfo {year} {2021})}\BibitemShut {NoStop}%
\bibitem [{\citenamefont {Bruzda}\ \emph {et~al.}(2023)\citenamefont {Bruzda},
  \citenamefont {Friedland},\ and\ \citenamefont
  {{\.Z}yczkowski}}]{bruzda2023rank}%
  \BibitemOpen
  \bibfield  {author} {\bibinfo {author} {\bibfnamefont {W.}~\bibnamefont
  {Bruzda}}, \bibinfo {author} {\bibfnamefont {S.}~\bibnamefont {Friedland}},\
  and\ \bibinfo {author} {\bibfnamefont {K.}~\bibnamefont {{\.Z}yczkowski}},\
  }\bibfield  {title} {\bibinfo {title} {Rank of a tensor and quantum
  entanglement},\ }\href {https://doi.org/10.1080/03081087.2023.2211717}
  {\bibfield  {journal} {\bibinfo  {journal} {Linear Multilinear Algebra}\ ,\
  \bibinfo {pages} {1}} (\bibinfo {year} {2023})}\BibitemShut {NoStop}%
\bibitem [{\citenamefont {Eisert}\ and\ \citenamefont
  {Briegel}(2001)}]{eisert2001schmidt}%
  \BibitemOpen
  \bibfield  {author} {\bibinfo {author} {\bibfnamefont {J.}~\bibnamefont
  {Eisert}}\ and\ \bibinfo {author} {\bibfnamefont {H.~J.}\ \bibnamefont
  {Briegel}},\ }\bibfield  {title} {\bibinfo {title} {Schmidt measure as a tool
  for quantifying multiparticle entanglement},\ }\href
  {https://doi.org/10.1103/physreva.64.022306} {\bibfield  {journal} {\bibinfo
  {journal} {Phys. Rev. A}\ }\textbf {\bibinfo {volume} {64}},\ \bibinfo
  {pages} {022306} (\bibinfo {year} {2001})}\BibitemShut {NoStop}%
\bibitem [{\citenamefont {Johnston}\ \emph {et~al.}(2022)\citenamefont
  {Johnston}, \citenamefont {Lovitz},\ and\ \citenamefont
  {Vijayaraghavan}}]{johnston2022complete}%
  \BibitemOpen
  \bibfield  {author} {\bibinfo {author} {\bibfnamefont {N.}~\bibnamefont
  {Johnston}}, \bibinfo {author} {\bibfnamefont {B.}~\bibnamefont {Lovitz}},\
  and\ \bibinfo {author} {\bibfnamefont {A.}~\bibnamefont {Vijayaraghavan}},\
  }\bibfield  {title} {\bibinfo {title} {Complete hierarchy of linear systems
  for certifying quantum entanglement of subspaces},\ }\href
  {https://doi.org/10.1103/physreva.106.062443} {\bibfield  {journal} {\bibinfo
   {journal} {Phys. Rev. A}\ }\textbf {\bibinfo {volume} {106}},\ \bibinfo
  {pages} {062443} (\bibinfo {year} {2022})}\BibitemShut {NoStop}%
\bibitem [{\citenamefont {Demianowicz}\ and\ \citenamefont
  {Augusiak}(2019)}]{demianowicz2019entanglement}%
  \BibitemOpen
  \bibfield  {author} {\bibinfo {author} {\bibfnamefont {M.}~\bibnamefont
  {Demianowicz}}\ and\ \bibinfo {author} {\bibfnamefont {R.}~\bibnamefont
  {Augusiak}},\ }\bibfield  {title} {\bibinfo {title} {Entanglement of
  genuinely entangled subspaces and states: Exact, approximate, and numerical
  results},\ }\href {https://doi.org/10.1103/physreva.100.062318} {\bibfield
  {journal} {\bibinfo  {journal} {Phys. Rev. A}\ }\textbf {\bibinfo {volume}
  {100}},\ \bibinfo {pages} {062318} (\bibinfo {year} {2019})}\BibitemShut
  {NoStop}%
\bibitem [{\citenamefont {Wei}\ and\ \citenamefont
  {Goldbart}(2003)}]{wei2003geometric}%
  \BibitemOpen
  \bibfield  {author} {\bibinfo {author} {\bibfnamefont {T.-C.}\ \bibnamefont
  {Wei}}\ and\ \bibinfo {author} {\bibfnamefont {P.~M.}\ \bibnamefont
  {Goldbart}},\ }\bibfield  {title} {\bibinfo {title} {Geometric measure of
  entanglement and applications to bipartite and multipartite quantum states},\
  }\href {https://doi.org/10.1103/physreva.68.042307} {\bibfield  {journal}
  {\bibinfo  {journal} {Phys. Rev. A}\ }\textbf {\bibinfo {volume} {68}},\
  \bibinfo {pages} {042307} (\bibinfo {year} {2003})}\BibitemShut {NoStop}%
\bibitem [{\citenamefont {Liu}\ \emph {et~al.}(2023)\citenamefont {Liu},
  \citenamefont {He}, \citenamefont {Huber}, \citenamefont {G\"uhne},\ and\
  \citenamefont {Vitagliano}}]{PRXQuantum.4.020324}%
  \BibitemOpen
  \bibfield  {author} {\bibinfo {author} {\bibfnamefont {S.}~\bibnamefont
  {Liu}}, \bibinfo {author} {\bibfnamefont {Q.}~\bibnamefont {He}}, \bibinfo
  {author} {\bibfnamefont {M.}~\bibnamefont {Huber}}, \bibinfo {author}
  {\bibfnamefont {O.}~\bibnamefont {G\"uhne}},\ and\ \bibinfo {author}
  {\bibfnamefont {G.}~\bibnamefont {Vitagliano}},\ }\bibfield  {title}
  {\bibinfo {title} {Characterizing entanglement dimensionality from randomized
  measurements},\ }\href {https://doi.org/10.1103/PRXQuantum.4.020324}
  {\bibfield  {journal} {\bibinfo  {journal} {PRX Quantum}\ }\textbf {\bibinfo
  {volume} {4}},\ \bibinfo {pages} {020324} (\bibinfo {year}
  {2023})}\BibitemShut {NoStop}%
\bibitem [{\citenamefont {Nape}\ \emph {et~al.}(2021)\citenamefont {Nape},
  \citenamefont {Rodr{\'\i}guez-Fajardo}, \citenamefont {Zhu}, \citenamefont
  {Huang}, \citenamefont {Leach},\ and\ \citenamefont
  {Forbes}}]{nape2021measuring}%
  \BibitemOpen
  \bibfield  {author} {\bibinfo {author} {\bibfnamefont {I.}~\bibnamefont
  {Nape}}, \bibinfo {author} {\bibfnamefont {V.}~\bibnamefont
  {Rodr{\'\i}guez-Fajardo}}, \bibinfo {author} {\bibfnamefont {F.}~\bibnamefont
  {Zhu}}, \bibinfo {author} {\bibfnamefont {H.-C.}\ \bibnamefont {Huang}},
  \bibinfo {author} {\bibfnamefont {J.}~\bibnamefont {Leach}},\ and\ \bibinfo
  {author} {\bibfnamefont {A.}~\bibnamefont {Forbes}},\ }\bibfield  {title}
  {\bibinfo {title} {Measuring dimensionality and purity of high-dimensional
  entangled states},\ }\href@noop {} {\bibfield  {journal} {\bibinfo  {journal}
  {Nature Communications}\ }\textbf {\bibinfo {volume} {12}},\ \bibinfo {pages}
  {5159} (\bibinfo {year} {2021})}\BibitemShut {NoStop}%
\bibitem [{\citenamefont {Ekert}\ and\ \citenamefont
  {Knight}(1995)}]{ekert1995entangled}%
  \BibitemOpen
  \bibfield  {author} {\bibinfo {author} {\bibfnamefont {A.}~\bibnamefont
  {Ekert}}\ and\ \bibinfo {author} {\bibfnamefont {P.~L.}\ \bibnamefont
  {Knight}},\ }\bibfield  {title} {\bibinfo {title} {Entangled quantum systems
  and the schmidt decomposition},\ }\href {https://doi.org/10.1119/1.17904}
  {\bibfield  {journal} {\bibinfo  {journal} {Am. J. Phys.}\ }\textbf {\bibinfo
  {volume} {63}},\ \bibinfo {pages} {415} (\bibinfo {year} {1995})}\BibitemShut
  {NoStop}%
\bibitem [{\citenamefont {Buss}\ \emph {et~al.}(1999)\citenamefont {Buss},
  \citenamefont {Frandsen},\ and\ \citenamefont
  {Shallit}}]{buss1999computational}%
  \BibitemOpen
  \bibfield  {author} {\bibinfo {author} {\bibfnamefont {J.~F.}\ \bibnamefont
  {Buss}}, \bibinfo {author} {\bibfnamefont {G.~S.}\ \bibnamefont {Frandsen}},\
  and\ \bibinfo {author} {\bibfnamefont {J.~O.}\ \bibnamefont {Shallit}},\
  }\bibfield  {title} {\bibinfo {title} {The computational complexity of some
  problems of linear algebra},\ }\href {https://doi.org/10.1006/jcss.1998.1608}
  {\bibfield  {journal} {\bibinfo  {journal} {J. Comput. Syst. Sci.}\ }\textbf
  {\bibinfo {volume} {58}},\ \bibinfo {pages} {572} (\bibinfo {year}
  {1999})}\BibitemShut {NoStop}%
\bibitem [{\citenamefont {Horodecki}(1997)}]{horodecki1997separability}%
  \BibitemOpen
  \bibfield  {author} {\bibinfo {author} {\bibfnamefont {P.}~\bibnamefont
  {Horodecki}},\ }\bibfield  {title} {\bibinfo {title} {Separability criterion
  and inseparable mixed states with positive partial transposition},\ }\href
  {https://doi.org/10.1016/s0375-9601(97)00416-7} {\bibfield  {journal}
  {\bibinfo  {journal} {Phys. Lett. A}\ }\textbf {\bibinfo {volume} {232}},\
  \bibinfo {pages} {333} (\bibinfo {year} {1997})}\BibitemShut {NoStop}%
\bibitem [{\citenamefont {Augusiak}\ \emph {et~al.}(2011)\citenamefont
  {Augusiak}, \citenamefont {Tura},\ and\ \citenamefont
  {Lewenstein}}]{augusiak2011note}%
  \BibitemOpen
  \bibfield  {author} {\bibinfo {author} {\bibfnamefont {R.}~\bibnamefont
  {Augusiak}}, \bibinfo {author} {\bibfnamefont {J.}~\bibnamefont {Tura}},\
  and\ \bibinfo {author} {\bibfnamefont {M.}~\bibnamefont {Lewenstein}},\
  }\bibfield  {title} {\bibinfo {title} {A note on the optimality of
  decomposable entanglement witnesses and completely entangled subspaces},\
  }\href {https://doi.org/10.1088/1751-8113/44/21/212001} {\bibfield  {journal}
  {\bibinfo  {journal} {J. Phys. A: Math. Theor.}\ }\textbf {\bibinfo {volume}
  {44}},\ \bibinfo {pages} {212001} (\bibinfo {year} {2011})}\BibitemShut
  {NoStop}%
\bibitem [{\citenamefont {Chru{\'s}ci{\'n}ski}\ and\ \citenamefont
  {Sarbicki}(2014)}]{chruscinski2014entanglement}%
  \BibitemOpen
  \bibfield  {author} {\bibinfo {author} {\bibfnamefont {D.}~\bibnamefont
  {Chru{\'s}ci{\'n}ski}}\ and\ \bibinfo {author} {\bibfnamefont
  {G.}~\bibnamefont {Sarbicki}},\ }\bibfield  {title} {\bibinfo {title}
  {Entanglement witnesses: construction, analysis and classification},\ }\href
  {https://doi.org/10.1088/1751-8113/47/48/483001} {\bibfield  {journal}
  {\bibinfo  {journal} {J. Phys. A: Math. Theor.}\ }\textbf {\bibinfo {volume}
  {47}},\ \bibinfo {pages} {483001} (\bibinfo {year} {2014})}\BibitemShut
  {NoStop}%
\bibitem [{\citenamefont {Gour}\ and\ \citenamefont
  {Wallach}(2007)}]{gour2007entanglement}%
  \BibitemOpen
  \bibfield  {author} {\bibinfo {author} {\bibfnamefont {G.}~\bibnamefont
  {Gour}}\ and\ \bibinfo {author} {\bibfnamefont {N.~R.}\ \bibnamefont
  {Wallach}},\ }\bibfield  {title} {\bibinfo {title} {Entanglement of subspaces
  and error-correcting codes},\ }\href
  {https://doi.org/10.1103/physreva.76.042309} {\bibfield  {journal} {\bibinfo
  {journal} {Phys. Rev. A}\ }\textbf {\bibinfo {volume} {76}},\ \bibinfo
  {pages} {042309} (\bibinfo {year} {2007})}\BibitemShut {NoStop}%
\bibitem [{\citenamefont {Huber}\ and\ \citenamefont
  {Grassl}(2020)}]{huber2020quantum}%
  \BibitemOpen
  \bibfield  {author} {\bibinfo {author} {\bibfnamefont {F.}~\bibnamefont
  {Huber}}\ and\ \bibinfo {author} {\bibfnamefont {M.}~\bibnamefont {Grassl}},\
  }\bibfield  {title} {\bibinfo {title} {Quantum codes of maximal distance and
  highly entangled subspaces},\ }\href
  {https://doi.org/10.22331/q-2020-06-18-284} {\bibfield  {journal} {\bibinfo
  {journal} {Quantum}\ }\textbf {\bibinfo {volume} {4}},\ \bibinfo {pages}
  {284} (\bibinfo {year} {2020})}\BibitemShut {NoStop}%
\bibitem [{\citenamefont {Horodecki}\ and\ \citenamefont
  {Piani}(2012)}]{horodecki2012quantum}%
  \BibitemOpen
  \bibfield  {author} {\bibinfo {author} {\bibfnamefont {M.}~\bibnamefont
  {Horodecki}}\ and\ \bibinfo {author} {\bibfnamefont {M.}~\bibnamefont
  {Piani}},\ }\bibfield  {title} {\bibinfo {title} {On quantum advantage in
  dense coding},\ }\href {https://doi.org/10.1088/1751-8113/45/10/105306}
  {\bibfield  {journal} {\bibinfo  {journal} {J. Phys. A: Math. Theor.}\
  }\textbf {\bibinfo {volume} {45}},\ \bibinfo {pages} {105306} (\bibinfo
  {year} {2012})}\BibitemShut {NoStop}%
\bibitem [{\citenamefont {Walter}\ \emph {et~al.}(2016)\citenamefont {Walter},
  \citenamefont {Gross},\ and\ \citenamefont
  {Eisert}}]{walter2016multipartite}%
  \BibitemOpen
  \bibfield  {author} {\bibinfo {author} {\bibfnamefont {M.}~\bibnamefont
  {Walter}}, \bibinfo {author} {\bibfnamefont {D.}~\bibnamefont {Gross}},\ and\
  \bibinfo {author} {\bibfnamefont {J.}~\bibnamefont {Eisert}},\ }\bibfield
  {title} {\bibinfo {title} {Multipartite entanglement},\ }\href
  {https://doi.org/10.1002/9783527805785.ch14} {\bibfield  {journal} {\bibinfo
  {journal} {Quantum Information}\ ,\ \bibinfo {pages} {293}} (\bibinfo {year}
  {2016})}\BibitemShut {NoStop}%
\bibitem [{\citenamefont {Dirac}(1939)}]{dirac1939new}%
  \BibitemOpen
  \bibfield  {author} {\bibinfo {author} {\bibfnamefont {P.~A.~M.}\
  \bibnamefont {Dirac}},\ }\bibfield  {title} {\bibinfo {title} {A new notation
  for quantum mechanics},\ }\href {https://doi.org/10.1017/S0305004100021162}
  {\bibfield  {journal} {\bibinfo  {journal} {Math. Proc. Cambridge Philos.
  Soc.}\ }\textbf {\bibinfo {volume} {35}},\ \bibinfo {pages} {416} (\bibinfo
  {year} {1939})}\BibitemShut {NoStop}%
\bibitem [{\citenamefont {H{\aa}stad}(1990)}]{haastad1990tensor}%
  \BibitemOpen
  \bibfield  {author} {\bibinfo {author} {\bibfnamefont {J.}~\bibnamefont
  {H{\aa}stad}},\ }\bibfield  {title} {\bibinfo {title} {Tensor rank is
  np-complete},\ }\href {https://doi.org/10.1016/0196-6774(90)90014-6}
  {\bibfield  {journal} {\bibinfo  {journal} {Journal of Algorithms}\ }\textbf
  {\bibinfo {volume} {11}},\ \bibinfo {pages} {644} (\bibinfo {year}
  {1990})}\BibitemShut {NoStop}%
\bibitem [{\citenamefont {Hillar}\ and\ \citenamefont
  {Lim}(2013)}]{hillar2013most}%
  \BibitemOpen
  \bibfield  {author} {\bibinfo {author} {\bibfnamefont {C.~J.}\ \bibnamefont
  {Hillar}}\ and\ \bibinfo {author} {\bibfnamefont {L.-H.}\ \bibnamefont
  {Lim}},\ }\bibfield  {title} {\bibinfo {title} {Most tensor problems are
  np-hard},\ }\href {https://doi.org/10.1145/2512329} {\bibfield  {journal}
  {\bibinfo  {journal} {J. ACM}\ }\textbf {\bibinfo {volume} {60}},\ \bibinfo
  {pages} {1} (\bibinfo {year} {2013})}\BibitemShut {NoStop}%
\bibitem [{\citenamefont {Lickteig}(1984)}]{lickteig1984note}%
  \BibitemOpen
  \bibfield  {author} {\bibinfo {author} {\bibfnamefont {T.}~\bibnamefont
  {Lickteig}},\ }\bibfield  {title} {\bibinfo {title} {A note on border rank},\
  }\href {https://doi.org/10.1016/0020-0190(84)90023-1} {\bibfield  {journal}
  {\bibinfo  {journal} {Inf. Process. Lett.}\ }\textbf {\bibinfo {volume}
  {18}},\ \bibinfo {pages} {173} (\bibinfo {year} {1984})}\BibitemShut
  {NoStop}%
\bibitem [{\citenamefont {Landsberg}\ and\ \citenamefont
  {Teitler}(2010)}]{landsberg2010ranks}%
  \BibitemOpen
  \bibfield  {author} {\bibinfo {author} {\bibfnamefont {J.~M.}\ \bibnamefont
  {Landsberg}}\ and\ \bibinfo {author} {\bibfnamefont {Z.}~\bibnamefont
  {Teitler}},\ }\bibfield  {title} {\bibinfo {title} {On the ranks and border
  ranks of symmetric tensors},\ }\href
  {https://doi.org/10.1007/s10208-009-9055-3} {\bibfield  {journal} {\bibinfo
  {journal} {Found. Comput. Math.}\ }\textbf {\bibinfo {volume} {10}},\
  \bibinfo {pages} {339} (\bibinfo {year} {2010})}\BibitemShut {NoStop}%
\bibitem [{\citenamefont {Erhard}\ \emph {et~al.}(2020)\citenamefont {Erhard},
  \citenamefont {Krenn},\ and\ \citenamefont {Zeilinger}}]{erhard2020advances}%
  \BibitemOpen
  \bibfield  {author} {\bibinfo {author} {\bibfnamefont {M.}~\bibnamefont
  {Erhard}}, \bibinfo {author} {\bibfnamefont {M.}~\bibnamefont {Krenn}},\ and\
  \bibinfo {author} {\bibfnamefont {A.}~\bibnamefont {Zeilinger}},\ }\bibfield
  {title} {\bibinfo {title} {Advances in high-dimensional quantum
  entanglement},\ }\href@noop {} {\bibfield  {journal} {\bibinfo  {journal}
  {Nature Reviews Physics}\ }\textbf {\bibinfo {volume} {2}},\ \bibinfo {pages}
  {365} (\bibinfo {year} {2020})}\BibitemShut {NoStop}%
\bibitem [{\citenamefont {Kong}\ \emph {et~al.}(2023)\citenamefont {Kong},
  \citenamefont {Sun}, \citenamefont {Zhang}, \citenamefont {Zhang},\ and\
  \citenamefont {Zhang}}]{kong2023high}%
  \BibitemOpen
  \bibfield  {author} {\bibinfo {author} {\bibfnamefont {L.-J.}\ \bibnamefont
  {Kong}}, \bibinfo {author} {\bibfnamefont {Y.}~\bibnamefont {Sun}}, \bibinfo
  {author} {\bibfnamefont {F.}~\bibnamefont {Zhang}}, \bibinfo {author}
  {\bibfnamefont {J.}~\bibnamefont {Zhang}},\ and\ \bibinfo {author}
  {\bibfnamefont {X.}~\bibnamefont {Zhang}},\ }\bibfield  {title} {\bibinfo
  {title} {High-dimensional entanglement-enabled holography},\ }\href@noop {}
  {\bibfield  {journal} {\bibinfo  {journal} {Physical Review Letters}\
  }\textbf {\bibinfo {volume} {130}},\ \bibinfo {pages} {053602} (\bibinfo
  {year} {2023})}\BibitemShut {NoStop}%
\bibitem [{\citenamefont {Cozzolino}\ \emph {et~al.}(2019)\citenamefont
  {Cozzolino}, \citenamefont {Da~Lio}, \citenamefont {Bacco},\ and\
  \citenamefont {Oxenl{\o}we}}]{cozzolino2019high}%
  \BibitemOpen
  \bibfield  {author} {\bibinfo {author} {\bibfnamefont {D.}~\bibnamefont
  {Cozzolino}}, \bibinfo {author} {\bibfnamefont {B.}~\bibnamefont {Da~Lio}},
  \bibinfo {author} {\bibfnamefont {D.}~\bibnamefont {Bacco}},\ and\ \bibinfo
  {author} {\bibfnamefont {L.~K.}\ \bibnamefont {Oxenl{\o}we}},\ }\bibfield
  {title} {\bibinfo {title} {High-dimensional quantum communication: benefits,
  progress, and future challenges},\ }\href@noop {} {\bibfield  {journal}
  {\bibinfo  {journal} {Advanced Quantum Technologies}\ }\textbf {\bibinfo
  {volume} {2}},\ \bibinfo {pages} {1900038} (\bibinfo {year}
  {2019})}\BibitemShut {NoStop}%
\bibitem [{\citenamefont {Etcheverry}\ \emph {et~al.}(2013)\citenamefont
  {Etcheverry}, \citenamefont {Ca{\~n}as}, \citenamefont {G{\'o}mez},
  \citenamefont {Nogueira}, \citenamefont {Saavedra}, \citenamefont {Xavier},\
  and\ \citenamefont {Lima}}]{etcheverry2013quantum}%
  \BibitemOpen
  \bibfield  {author} {\bibinfo {author} {\bibfnamefont {S.}~\bibnamefont
  {Etcheverry}}, \bibinfo {author} {\bibfnamefont {G.}~\bibnamefont
  {Ca{\~n}as}}, \bibinfo {author} {\bibfnamefont {E.}~\bibnamefont
  {G{\'o}mez}}, \bibinfo {author} {\bibfnamefont {W.}~\bibnamefont {Nogueira}},
  \bibinfo {author} {\bibfnamefont {C.}~\bibnamefont {Saavedra}}, \bibinfo
  {author} {\bibfnamefont {G.}~\bibnamefont {Xavier}},\ and\ \bibinfo {author}
  {\bibfnamefont {G.}~\bibnamefont {Lima}},\ }\bibfield  {title} {\bibinfo
  {title} {Quantum key distribution session with 16-dimensional photonic
  states},\ }\href@noop {} {\bibfield  {journal} {\bibinfo  {journal}
  {Scientific reports}\ }\textbf {\bibinfo {volume} {3}},\ \bibinfo {pages}
  {2316} (\bibinfo {year} {2013})}\BibitemShut {NoStop}%
\bibitem [{\citenamefont {Strassen}\ \emph {et~al.}(1969)\citenamefont
  {Strassen} \emph {et~al.}}]{strassen1969gaussian}%
  \BibitemOpen
  \bibfield  {author} {\bibinfo {author} {\bibfnamefont {V.}~\bibnamefont
  {Strassen}} \emph {et~al.},\ }\bibfield  {title} {\bibinfo {title} {Gaussian
  elimination is not optimal},\ }\href {https://doi.org/10.1007/bf02165411}
  {\bibfield  {journal} {\bibinfo  {journal} {Numer. Math.}\ }\textbf {\bibinfo
  {volume} {13}},\ \bibinfo {pages} {354} (\bibinfo {year} {1969})}\BibitemShut
  {NoStop}%
\bibitem [{\citenamefont {Bini}\ \emph {et~al.}(1979)\citenamefont {Bini},
  \citenamefont {Capovani}, \citenamefont {Romani},\ and\ \citenamefont
  {Lotti}}]{bini1979n2}%
  \BibitemOpen
  \bibfield  {author} {\bibinfo {author} {\bibfnamefont {D.}~\bibnamefont
  {Bini}}, \bibinfo {author} {\bibfnamefont {M.}~\bibnamefont {Capovani}},
  \bibinfo {author} {\bibfnamefont {F.}~\bibnamefont {Romani}},\ and\ \bibinfo
  {author} {\bibfnamefont {G.}~\bibnamefont {Lotti}},\ }\bibfield  {title}
  {\bibinfo {title} {O (n2. 7799) complexity for n$\times$ n approximate matrix
  multiplication},\ }\href {https://doi.org/10.1016/0020-0190(79)90113-3}
  {\bibfield  {journal} {\bibinfo  {journal} {Inf. Process. Lett.}\ }\textbf
  {\bibinfo {volume} {8}},\ \bibinfo {pages} {234} (\bibinfo {year}
  {1979})}\BibitemShut {NoStop}%
\bibitem [{\citenamefont {Bini}\ \emph {et~al.}(1980)\citenamefont {Bini},
  \citenamefont {Lotti},\ and\ \citenamefont {Romani}}]{bini1980approximate}%
  \BibitemOpen
  \bibfield  {author} {\bibinfo {author} {\bibfnamefont {D.}~\bibnamefont
  {Bini}}, \bibinfo {author} {\bibfnamefont {G.}~\bibnamefont {Lotti}},\ and\
  \bibinfo {author} {\bibfnamefont {F.}~\bibnamefont {Romani}},\ }\bibfield
  {title} {\bibinfo {title} {Approximate solutions for the bilinear form
  computational problem},\ }\href {https://doi.org/10.1137/0209053} {\bibfield
  {journal} {\bibinfo  {journal} {SIAM J. Comput.}\ }\textbf {\bibinfo {volume}
  {9}},\ \bibinfo {pages} {692} (\bibinfo {year} {1980})}\BibitemShut {NoStop}%
\bibitem [{\citenamefont {Bennett}\ \emph {et~al.}(1999)\citenamefont
  {Bennett}, \citenamefont {DiVincenzo}, \citenamefont {Mor}, \citenamefont
  {Shor}, \citenamefont {Smolin},\ and\ \citenamefont
  {Terhal}}]{bennett1999unextendible}%
  \BibitemOpen
  \bibfield  {author} {\bibinfo {author} {\bibfnamefont {C.~H.}\ \bibnamefont
  {Bennett}}, \bibinfo {author} {\bibfnamefont {D.~P.}\ \bibnamefont
  {DiVincenzo}}, \bibinfo {author} {\bibfnamefont {T.}~\bibnamefont {Mor}},
  \bibinfo {author} {\bibfnamefont {P.~W.}\ \bibnamefont {Shor}}, \bibinfo
  {author} {\bibfnamefont {J.~A.}\ \bibnamefont {Smolin}},\ and\ \bibinfo
  {author} {\bibfnamefont {B.~M.}\ \bibnamefont {Terhal}},\ }\bibfield  {title}
  {\bibinfo {title} {Unextendible product bases and bound entanglement},\
  }\href {https://doi.org/10.1103/physrevlett.82.5385} {\bibfield  {journal}
  {\bibinfo  {journal} {Phys. Rev. Lett.}\ }\textbf {\bibinfo {volume} {82}},\
  \bibinfo {pages} {5385} (\bibinfo {year} {1999})}\BibitemShut {NoStop}%
\bibitem [{\citenamefont {Demianowicz}\ and\ \citenamefont
  {Augusiak}(2018)}]{demianowicz2018unextendible}%
  \BibitemOpen
  \bibfield  {author} {\bibinfo {author} {\bibfnamefont {M.}~\bibnamefont
  {Demianowicz}}\ and\ \bibinfo {author} {\bibfnamefont {R.}~\bibnamefont
  {Augusiak}},\ }\bibfield  {title} {\bibinfo {title} {From unextendible
  product bases to genuinely entangled subspaces},\ }\href
  {https://doi.org/10.1103/physreva.98.012313} {\bibfield  {journal} {\bibinfo
  {journal} {Phys. Rev. A}\ }\textbf {\bibinfo {volume} {98}},\ \bibinfo
  {pages} {012313} (\bibinfo {year} {2018})}\BibitemShut {NoStop}%
\bibitem [{\citenamefont {Agrawal}\ \emph {et~al.}(2019)\citenamefont
  {Agrawal}, \citenamefont {Halder},\ and\ \citenamefont
  {Banik}}]{agrawal2019genuinely}%
  \BibitemOpen
  \bibfield  {author} {\bibinfo {author} {\bibfnamefont {S.}~\bibnamefont
  {Agrawal}}, \bibinfo {author} {\bibfnamefont {S.}~\bibnamefont {Halder}},\
  and\ \bibinfo {author} {\bibfnamefont {M.}~\bibnamefont {Banik}},\ }\bibfield
   {title} {\bibinfo {title} {Genuinely entangled subspace with
  all-encompassing distillable entanglement across every bipartition},\ }\href
  {https://doi.org/10.1103/physreva.99.032335} {\bibfield  {journal} {\bibinfo
  {journal} {Phys. Rev. A}\ }\textbf {\bibinfo {volume} {99}},\ \bibinfo
  {pages} {032335} (\bibinfo {year} {2019})}\BibitemShut {NoStop}%
\bibitem [{\citenamefont {Lovitz}\ and\ \citenamefont
  {Johnston}(2022)}]{lovitz2022entangled}%
  \BibitemOpen
  \bibfield  {author} {\bibinfo {author} {\bibfnamefont {B.}~\bibnamefont
  {Lovitz}}\ and\ \bibinfo {author} {\bibfnamefont {N.}~\bibnamefont
  {Johnston}},\ }\bibfield  {title} {\bibinfo {title} {Entangled subspaces and
  generic local state discrimination with pre-shared entanglement},\ }\href
  {https://doi.org/10.22331/q-2022-07-07-760} {\bibfield  {journal} {\bibinfo
  {journal} {Quantum}\ }\textbf {\bibinfo {volume} {6}},\ \bibinfo {pages}
  {760} (\bibinfo {year} {2022})}\BibitemShut {NoStop}%
\bibitem [{\citenamefont {Shenoy}\ and\ \citenamefont
  {Srikanth}(2019)}]{shenoy2019maximally}%
  \BibitemOpen
  \bibfield  {author} {\bibinfo {author} {\bibfnamefont {A.~H.}\ \bibnamefont
  {Shenoy}}\ and\ \bibinfo {author} {\bibfnamefont {R.}~\bibnamefont
  {Srikanth}},\ }\bibfield  {title} {\bibinfo {title} {Maximally nonlocal
  subspaces},\ }\href {https://doi.org/10.1088/1751-8121/ab0046} {\bibfield
  {journal} {\bibinfo  {journal} {J. Phys. A: Math. Theor.}\ }\textbf {\bibinfo
  {volume} {52}},\ \bibinfo {pages} {095302} (\bibinfo {year}
  {2019})}\BibitemShut {NoStop}%
\bibitem [{\citenamefont {Cavalcanti}(2006)}]{cavalcanti2006connecting}%
  \BibitemOpen
  \bibfield  {author} {\bibinfo {author} {\bibfnamefont {D.}~\bibnamefont
  {Cavalcanti}},\ }\bibfield  {title} {\bibinfo {title} {Connecting the
  generalized robustness and the geometric measure of entanglement},\ }\href
  {https://doi.org/10.1103/physreva.73.044302} {\bibfield  {journal} {\bibinfo
  {journal} {Phys. Rev. A}\ }\textbf {\bibinfo {volume} {73}},\ \bibinfo
  {pages} {044302} (\bibinfo {year} {2006})}\BibitemShut {NoStop}%
\bibitem [{\citenamefont {Ou}\ and\ \citenamefont {Fan}(2007)}]{ou2007bounds}%
  \BibitemOpen
  \bibfield  {author} {\bibinfo {author} {\bibfnamefont {Y.-C.}\ \bibnamefont
  {Ou}}\ and\ \bibinfo {author} {\bibfnamefont {H.}~\bibnamefont {Fan}},\
  }\bibfield  {title} {\bibinfo {title} {Bounds on negativity of
  superpositions},\ }\href {https://doi.org/10.1103/physreva.76.022320}
  {\bibfield  {journal} {\bibinfo  {journal} {Phys. Rev. A}\ }\textbf {\bibinfo
  {volume} {76}},\ \bibinfo {pages} {022320} (\bibinfo {year}
  {2007})}\BibitemShut {NoStop}%
\bibitem [{\citenamefont {Niset}\ and\ \citenamefont
  {Cerf}(2007)}]{niset2007tight}%
  \BibitemOpen
  \bibfield  {author} {\bibinfo {author} {\bibfnamefont {J.}~\bibnamefont
  {Niset}}\ and\ \bibinfo {author} {\bibfnamefont {N.~J.}\ \bibnamefont
  {Cerf}},\ }\bibfield  {title} {\bibinfo {title} {Tight bounds on the
  concurrence of quantum superpositions},\ }\href
  {https://doi.org/10.1103/physreva.76.042328} {\bibfield  {journal} {\bibinfo
  {journal} {Phys. Rev. A}\ }\textbf {\bibinfo {volume} {76}},\ \bibinfo
  {pages} {042328} (\bibinfo {year} {2007})}\BibitemShut {NoStop}%
\bibitem [{\citenamefont {Song}\ \emph {et~al.}(2007)\citenamefont {Song},
  \citenamefont {Liu},\ and\ \citenamefont {Chen}}]{song2007bounds}%
  \BibitemOpen
  \bibfield  {author} {\bibinfo {author} {\bibfnamefont {W.}~\bibnamefont
  {Song}}, \bibinfo {author} {\bibfnamefont {N.-L.}\ \bibnamefont {Liu}},\ and\
  \bibinfo {author} {\bibfnamefont {Z.-B.}\ \bibnamefont {Chen}},\ }\bibfield
  {title} {\bibinfo {title} {Bounds on the multipartite entanglement of
  superpositions},\ }\href {https://doi.org/10.1103/physreva.76.054303}
  {\bibfield  {journal} {\bibinfo  {journal} {Phys. Rev. A}\ }\textbf {\bibinfo
  {volume} {76}},\ \bibinfo {pages} {054303} (\bibinfo {year}
  {2007})}\BibitemShut {NoStop}%
\bibitem [{\citenamefont {Xiang}\ \emph {et~al.}(2008)\citenamefont {Xiang},
  \citenamefont {Xiong},\ and\ \citenamefont {Hong}}]{xiang2008bound}%
  \BibitemOpen
  \bibfield  {author} {\bibinfo {author} {\bibfnamefont {Y.}~\bibnamefont
  {Xiang}}, \bibinfo {author} {\bibfnamefont {S.-J.}\ \bibnamefont {Xiong}},\
  and\ \bibinfo {author} {\bibfnamefont {F.-Y.}\ \bibnamefont {Hong}},\
  }\bibfield  {title} {\bibinfo {title} {The bound of entanglement of
  superpositions with more than two components},\ }\href
  {https://doi.org/10.1140/epjd/e2008-00022-6} {\bibfield  {journal} {\bibinfo
  {journal} {Eur. Phys. J. D}\ }\textbf {\bibinfo {volume} {47}},\ \bibinfo
  {pages} {257} (\bibinfo {year} {2008})}\BibitemShut {NoStop}%
\bibitem [{\citenamefont {Akhtarshenas}(2011)}]{akhtarshenas2011concurrence}%
  \BibitemOpen
  \bibfield  {author} {\bibinfo {author} {\bibfnamefont {S.~J.}\ \bibnamefont
  {Akhtarshenas}},\ }\bibfield  {title} {\bibinfo {title} {Concurrence of
  superpositions of many states},\ }\href
  {https://doi.org/10.1103/physreva.83.042306} {\bibfield  {journal} {\bibinfo
  {journal} {Phys. Rev. A}\ }\textbf {\bibinfo {volume} {83}},\ \bibinfo
  {pages} {042306} (\bibinfo {year} {2011})}\BibitemShut {NoStop}%
\bibitem [{\citenamefont {Ma}\ \emph {et~al.}(2012)\citenamefont {Ma},
  \citenamefont {Chen}, \citenamefont {Han}, \citenamefont {Fei},\ and\
  \citenamefont {Severini}}]{ma2012improved}%
  \BibitemOpen
  \bibfield  {author} {\bibinfo {author} {\bibfnamefont {Z.}~\bibnamefont
  {Ma}}, \bibinfo {author} {\bibfnamefont {Z.}~\bibnamefont {Chen}}, \bibinfo
  {author} {\bibfnamefont {S.}~\bibnamefont {Han}}, \bibinfo {author}
  {\bibfnamefont {S.}~\bibnamefont {Fei}},\ and\ \bibinfo {author}
  {\bibfnamefont {S.}~\bibnamefont {Severini}},\ }\bibfield  {title} {\bibinfo
  {title} {Improved bounds on negativity of superpositions},\ }\href
  {https://doi.org/10.26421/QIC12.11-12-6} {\bibfield  {journal} {\bibinfo
  {journal} {Quantum Inf. Comput.}\ }\textbf {\bibinfo {volume} {12}},\
  \bibinfo {pages} {983} (\bibinfo {year} {2012})}\BibitemShut {NoStop}%
\bibitem [{\citenamefont {Linden}\ \emph {et~al.}(2006)\citenamefont {Linden},
  \citenamefont {Popescu},\ and\ \citenamefont
  {Smolin}}]{linden2006entanglement}%
  \BibitemOpen
  \bibfield  {author} {\bibinfo {author} {\bibfnamefont {N.}~\bibnamefont
  {Linden}}, \bibinfo {author} {\bibfnamefont {S.}~\bibnamefont {Popescu}},\
  and\ \bibinfo {author} {\bibfnamefont {J.~A.}\ \bibnamefont {Smolin}},\
  }\bibfield  {title} {\bibinfo {title} {Entanglement of superpositions},\
  }\href {https://doi.org/10.1103/physrevlett.97.100502} {\bibfield  {journal}
  {\bibinfo  {journal} {Phys. Rev. Lett.}\ }\textbf {\bibinfo {volume} {97}},\
  \bibinfo {pages} {100502} (\bibinfo {year} {2006})}\BibitemShut {NoStop}%
\bibitem [{\citenamefont {Zhang}\ \emph {et~al.}(2020)\citenamefont {Zhang},
  \citenamefont {Dai}, \citenamefont {Dong},\ and\ \citenamefont
  {Zhang}}]{zhang2020numerical}%
  \BibitemOpen
  \bibfield  {author} {\bibinfo {author} {\bibfnamefont {Z.}~\bibnamefont
  {Zhang}}, \bibinfo {author} {\bibfnamefont {Y.}~\bibnamefont {Dai}}, \bibinfo
  {author} {\bibfnamefont {Y.-L.}\ \bibnamefont {Dong}},\ and\ \bibinfo
  {author} {\bibfnamefont {C.}~\bibnamefont {Zhang}},\ }\bibfield  {title}
  {\bibinfo {title} {Numerical and analytical results for geometric measure of
  coherence and geometric measure of entanglement},\ }\href
  {https://doi.org/10.1038/s41598-020-68979-z} {\bibfield  {journal} {\bibinfo
  {journal} {Sci. Rep.}\ }\textbf {\bibinfo {volume} {10}},\ \bibinfo {pages}
  {12122} (\bibinfo {year} {2020})}\BibitemShut {NoStop}%
\bibitem [{\citenamefont {Doherty}\ \emph {et~al.}(2004)\citenamefont
  {Doherty}, \citenamefont {Parrilo},\ and\ \citenamefont
  {Spedalieri}}]{doherty2004complete}%
  \BibitemOpen
  \bibfield  {author} {\bibinfo {author} {\bibfnamefont {A.~C.}\ \bibnamefont
  {Doherty}}, \bibinfo {author} {\bibfnamefont {P.~A.}\ \bibnamefont
  {Parrilo}},\ and\ \bibinfo {author} {\bibfnamefont {F.~M.}\ \bibnamefont
  {Spedalieri}},\ }\bibfield  {title} {\bibinfo {title} {Complete family of
  separability criteria},\ }\href {https://doi.org/10.1103/physreva.69.022308}
  {\bibfield  {journal} {\bibinfo  {journal} {Phys. Rev. A}\ }\textbf {\bibinfo
  {volume} {69}},\ \bibinfo {pages} {022308} (\bibinfo {year}
  {2004})}\BibitemShut {NoStop}%
\bibitem [{\citenamefont {Horodecki}\ \emph {et~al.}(1998)\citenamefont
  {Horodecki}, \citenamefont {Horodecki},\ and\ \citenamefont
  {Horodecki}}]{horodecki1998mixed}%
  \BibitemOpen
  \bibfield  {author} {\bibinfo {author} {\bibfnamefont {M.}~\bibnamefont
  {Horodecki}}, \bibinfo {author} {\bibfnamefont {P.}~\bibnamefont
  {Horodecki}},\ and\ \bibinfo {author} {\bibfnamefont {R.}~\bibnamefont
  {Horodecki}},\ }\bibfield  {title} {\bibinfo {title} {Mixed-state
  entanglement and distillation: Is there a “bound” entanglement in
  nature?},\ }\href {https://doi.org/10.1103/physrevlett.80.5239} {\bibfield
  {journal} {\bibinfo  {journal} {Phys. Rev. Lett.}\ }\textbf {\bibinfo
  {volume} {80}},\ \bibinfo {pages} {5239} (\bibinfo {year}
  {1998})}\BibitemShut {NoStop}%
\bibitem [{\citenamefont {Cubitt}\ \emph {et~al.}(2008)\citenamefont {Cubitt},
  \citenamefont {Montanaro},\ and\ \citenamefont
  {Winter}}]{cubitt2008dimension}%
  \BibitemOpen
  \bibfield  {author} {\bibinfo {author} {\bibfnamefont {T.}~\bibnamefont
  {Cubitt}}, \bibinfo {author} {\bibfnamefont {A.}~\bibnamefont {Montanaro}},\
  and\ \bibinfo {author} {\bibfnamefont {A.}~\bibnamefont {Winter}},\
  }\bibfield  {title} {\bibinfo {title} {On the dimension of subspaces with
  bounded schmidt rank},\ }\bibfield  {journal} {\bibinfo  {journal} {J. Math.
  Phys.}\ }\textbf {\bibinfo {volume} {49}},\ \href
  {https://doi.org/10.1063/1.2862998} {10.1063/1.2862998} (\bibinfo {year}
  {2008})\BibitemShut {NoStop}%
\bibitem [{\citenamefont {Blasone}\ \emph {et~al.}(2008)\citenamefont
  {Blasone}, \citenamefont {Dell'Anno}, \citenamefont {De~Siena},\ and\
  \citenamefont {Illuminati}}]{blasone2008hierarchies}%
  \BibitemOpen
  \bibfield  {author} {\bibinfo {author} {\bibfnamefont {M.}~\bibnamefont
  {Blasone}}, \bibinfo {author} {\bibfnamefont {F.}~\bibnamefont {Dell'Anno}},
  \bibinfo {author} {\bibfnamefont {S.}~\bibnamefont {De~Siena}},\ and\
  \bibinfo {author} {\bibfnamefont {F.}~\bibnamefont {Illuminati}},\ }\bibfield
   {title} {\bibinfo {title} {Hierarchies of geometric entanglement},\ }\href
  {https://doi.org/10.1103/physreva.77.062304} {\bibfield  {journal} {\bibinfo
  {journal} {Phys. Rev. A}\ }\textbf {\bibinfo {volume} {77}},\ \bibinfo
  {pages} {062304} (\bibinfo {year} {2008})}\BibitemShut {NoStop}%
\bibitem [{\citenamefont {Paszke}\ \emph {et~al.}(2019)\citenamefont {Paszke},
  \citenamefont {Gross}, \citenamefont {Massa}, \citenamefont {Lerer},
  \citenamefont {Bradbury}, \citenamefont {Chanan}, \citenamefont {Killeen},
  \citenamefont {Lin}, \citenamefont {Gimelshein}, \citenamefont {Antiga},
  \citenamefont {Desmaison}, \citenamefont {Kopf}, \citenamefont {Yang},
  \citenamefont {DeVito}, \citenamefont {Raison}, \citenamefont {Tejani},
  \citenamefont {Chilamkurthy}, \citenamefont {Steiner}, \citenamefont {Fang},
  \citenamefont {Bai},\ and\ \citenamefont {Chintala}}]{pytorch}%
  \BibitemOpen
  \bibfield  {author} {\bibinfo {author} {\bibfnamefont {A.}~\bibnamefont
  {Paszke}}, \bibinfo {author} {\bibfnamefont {S.}~\bibnamefont {Gross}},
  \bibinfo {author} {\bibfnamefont {F.}~\bibnamefont {Massa}}, \bibinfo
  {author} {\bibfnamefont {A.}~\bibnamefont {Lerer}}, \bibinfo {author}
  {\bibfnamefont {J.}~\bibnamefont {Bradbury}}, \bibinfo {author}
  {\bibfnamefont {G.}~\bibnamefont {Chanan}}, \bibinfo {author} {\bibfnamefont
  {T.}~\bibnamefont {Killeen}}, \bibinfo {author} {\bibfnamefont
  {Z.}~\bibnamefont {Lin}}, \bibinfo {author} {\bibfnamefont {N.}~\bibnamefont
  {Gimelshein}}, \bibinfo {author} {\bibfnamefont {L.}~\bibnamefont {Antiga}},
  \bibinfo {author} {\bibfnamefont {A.}~\bibnamefont {Desmaison}}, \bibinfo
  {author} {\bibfnamefont {A.}~\bibnamefont {Kopf}}, \bibinfo {author}
  {\bibfnamefont {E.}~\bibnamefont {Yang}}, \bibinfo {author} {\bibfnamefont
  {Z.}~\bibnamefont {DeVito}}, \bibinfo {author} {\bibfnamefont
  {M.}~\bibnamefont {Raison}}, \bibinfo {author} {\bibfnamefont
  {A.}~\bibnamefont {Tejani}}, \bibinfo {author} {\bibfnamefont
  {S.}~\bibnamefont {Chilamkurthy}}, \bibinfo {author} {\bibfnamefont
  {B.}~\bibnamefont {Steiner}}, \bibinfo {author} {\bibfnamefont
  {L.}~\bibnamefont {Fang}}, \bibinfo {author} {\bibfnamefont {J.}~\bibnamefont
  {Bai}},\ and\ \bibinfo {author} {\bibfnamefont {S.}~\bibnamefont
  {Chintala}},\ }\bibfield  {title} {\bibinfo {title} {{PyTorch: An Imperative
  Style, High-Performance Deep Learning Library}},\ }in\ \href
  {http://papers.neurips.cc/paper/9015-pytorch-an-imperative-style-high-performance-deep-learning-library.pdf}
  {\emph {\bibinfo {booktitle} {Advances in Neural Information Processing
  Systems 32}}},\ \bibinfo {editor} {edited by\ \bibinfo {editor}
  {\bibfnamefont {H.}~\bibnamefont {Wallach}}, \bibinfo {editor} {\bibfnamefont
  {H.}~\bibnamefont {Larochelle}}, \bibinfo {editor} {\bibfnamefont
  {A.}~\bibnamefont {Beygelzimer}}, \bibinfo {editor} {\bibfnamefont
  {F.}~\bibnamefont {d'Alché Buc}}, \bibinfo {editor} {\bibfnamefont
  {E.}~\bibnamefont {Fox}},\ and\ \bibinfo {editor} {\bibfnamefont
  {R.}~\bibnamefont {Garnett}}}\ (\bibinfo  {publisher} {Curran Associates,
  Inc.},\ \bibinfo {year} {2019})\ pp.\ \bibinfo {pages}
  {8024--8035}\BibitemShut {NoStop}%
\bibitem [{\citenamefont {Casado}(2019)}]{lezcanocasado2019trivializations}%
  \BibitemOpen
  \bibfield  {author} {\bibinfo {author} {\bibfnamefont {M.~L.}\ \bibnamefont
  {Casado}},\ }\href {http://arxiv.org/abs/1909.09501} {\bibinfo {title}
  {Trivializations for gradient-based optimization on manifolds}} (\bibinfo
  {year} {2019}),\ \Eprint {https://arxiv.org/abs/1909.09501} {1909.09501}
  \BibitemShut {NoStop}%
\bibitem [{\citenamefont {Hu}\ \emph {et~al.}(2020)\citenamefont {Hu},
  \citenamefont {Liu}, \citenamefont {Wen},\ and\ \citenamefont
  {Yuan}}]{hu2020brief}%
  \BibitemOpen
  \bibfield  {author} {\bibinfo {author} {\bibfnamefont {J.}~\bibnamefont
  {Hu}}, \bibinfo {author} {\bibfnamefont {X.}~\bibnamefont {Liu}}, \bibinfo
  {author} {\bibfnamefont {Z.-W.}\ \bibnamefont {Wen}},\ and\ \bibinfo {author}
  {\bibfnamefont {Y.-X.}\ \bibnamefont {Yuan}},\ }\bibfield  {title} {\bibinfo
  {title} {A brief introduction to manifold optimization},\ }\href@noop {}
  {\bibfield  {journal} {\bibinfo  {journal} {Journal of the Operations
  Research Society of China}\ }\textbf {\bibinfo {volume} {8}},\ \bibinfo
  {pages} {199} (\bibinfo {year} {2020})}\BibitemShut {NoStop}%
\bibitem [{\citenamefont {Absil}\ \emph {et~al.}(2008)\citenamefont {Absil},
  \citenamefont {Mahony},\ and\ \citenamefont
  {Sepulchre}}]{absil2008optimization}%
  \BibitemOpen
  \bibfield  {author} {\bibinfo {author} {\bibfnamefont {P.-A.}\ \bibnamefont
  {Absil}}, \bibinfo {author} {\bibfnamefont {R.}~\bibnamefont {Mahony}},\ and\
  \bibinfo {author} {\bibfnamefont {R.}~\bibnamefont {Sepulchre}},\ }\href@noop
  {} {\emph {\bibinfo {title} {Optimization algorithms on matrix manifolds}}}\
  (\bibinfo  {publisher} {Princeton University Press},\ \bibinfo {year}
  {2008})\BibitemShut {NoStop}%
\bibitem [{\citenamefont {Carteret}\ \emph {et~al.}(2000)\citenamefont
  {Carteret}, \citenamefont {Higuchi},\ and\ \citenamefont
  {Sudbery}}]{carteret2000multipartite}%
  \BibitemOpen
  \bibfield  {author} {\bibinfo {author} {\bibfnamefont {H.~A.}\ \bibnamefont
  {Carteret}}, \bibinfo {author} {\bibfnamefont {A.}~\bibnamefont {Higuchi}},\
  and\ \bibinfo {author} {\bibfnamefont {A.}~\bibnamefont {Sudbery}},\
  }\bibfield  {title} {\bibinfo {title} {Multipartite generalization of the
  schmidt decomposition},\ }\href {https://doi.org/10.1063/1.1319516}
  {\bibfield  {journal} {\bibinfo  {journal} {J. Math. Phys.}\ }\textbf
  {\bibinfo {volume} {41}},\ \bibinfo {pages} {7932} (\bibinfo {year}
  {2000})}\BibitemShut {NoStop}%
\bibitem [{\citenamefont {Chen}\ \emph {et~al.}(2010)\citenamefont {Chen},
  \citenamefont {Chitambar}, \citenamefont {Duan}, \citenamefont {Ji},\ and\
  \citenamefont {Winter}}]{chen2010tensor}%
  \BibitemOpen
  \bibfield  {author} {\bibinfo {author} {\bibfnamefont {L.}~\bibnamefont
  {Chen}}, \bibinfo {author} {\bibfnamefont {E.}~\bibnamefont {Chitambar}},
  \bibinfo {author} {\bibfnamefont {R.}~\bibnamefont {Duan}}, \bibinfo {author}
  {\bibfnamefont {Z.}~\bibnamefont {Ji}},\ and\ \bibinfo {author}
  {\bibfnamefont {A.}~\bibnamefont {Winter}},\ }\bibfield  {title} {\bibinfo
  {title} {Tensor rank and stochastic entanglement catalysis for multipartite
  pure states},\ }\href {https://doi.org/10.1103/physrevlett.105.200501}
  {\bibfield  {journal} {\bibinfo  {journal} {Phys. Rev. Lett.}\ }\textbf
  {\bibinfo {volume} {105}},\ \bibinfo {pages} {200501} (\bibinfo {year}
  {2010})}\BibitemShut {NoStop}%
\bibitem [{\citenamefont {De~Silva}\ and\ \citenamefont
  {Lim}(2008)}]{de2008tensor}%
  \BibitemOpen
  \bibfield  {author} {\bibinfo {author} {\bibfnamefont {V.}~\bibnamefont
  {De~Silva}}\ and\ \bibinfo {author} {\bibfnamefont {L.-H.}\ \bibnamefont
  {Lim}},\ }\bibfield  {title} {\bibinfo {title} {Tensor rank and the
  ill-posedness of the best low-rank approximation problem},\ }\href
  {https://doi.org/10.1137/06066518x} {\bibfield  {journal} {\bibinfo
  {journal} {SIAM J. Matrix Anal. Appl.}\ }\textbf {\bibinfo {volume} {30}},\
  \bibinfo {pages} {1084} (\bibinfo {year} {2008})}\BibitemShut {NoStop}%
\bibitem [{\citenamefont {Branciard}\ \emph {et~al.}(2010)\citenamefont
  {Branciard}, \citenamefont {Zhu}, \citenamefont {Chen},\ and\ \citenamefont
  {Scarani}}]{branciard2010evaluation}%
  \BibitemOpen
  \bibfield  {author} {\bibinfo {author} {\bibfnamefont {C.}~\bibnamefont
  {Branciard}}, \bibinfo {author} {\bibfnamefont {H.}~\bibnamefont {Zhu}},
  \bibinfo {author} {\bibfnamefont {L.}~\bibnamefont {Chen}},\ and\ \bibinfo
  {author} {\bibfnamefont {V.}~\bibnamefont {Scarani}},\ }\bibfield  {title}
  {\bibinfo {title} {Evaluation of two different entanglement measures on a
  bound entangled state},\ }\href {https://doi.org/10.1103/physreva.82.012327}
  {\bibfield  {journal} {\bibinfo  {journal} {Phys. Rev. A}\ }\textbf {\bibinfo
  {volume} {82}},\ \bibinfo {pages} {012327} (\bibinfo {year}
  {2010})}\BibitemShut {NoStop}%
\bibitem [{\citenamefont {Zheng}\ \emph {et~al.}(2015)\citenamefont {Zheng},
  \citenamefont {Yang}, \citenamefont {Liu}, \citenamefont {Liang},\ and\
  \citenamefont {Li}}]{softplus}%
  \BibitemOpen
  \bibfield  {author} {\bibinfo {author} {\bibfnamefont {H.}~\bibnamefont
  {Zheng}}, \bibinfo {author} {\bibfnamefont {Z.}~\bibnamefont {Yang}},
  \bibinfo {author} {\bibfnamefont {W.}~\bibnamefont {Liu}}, \bibinfo {author}
  {\bibfnamefont {J.}~\bibnamefont {Liang}},\ and\ \bibinfo {author}
  {\bibfnamefont {Y.}~\bibnamefont {Li}},\ }\bibfield  {title} {\bibinfo
  {title} {Improving deep neural networks using softplus units},\ }in\ \href
  {https://doi.org/10.1109/IJCNN.2015.7280459} {\emph {\bibinfo {booktitle}
  {2015 International Joint Conference on Neural Networks (IJCNN)}}}\ (\bibinfo
  {year} {2015})\ pp.\ \bibinfo {pages} {1--4}\BibitemShut {NoStop}%
\bibitem [{\citenamefont {Virtanen}\ \emph {et~al.}(2020)\citenamefont
  {Virtanen}, \citenamefont {Gommers}, \citenamefont {Oliphant}, \citenamefont
  {Haberland}, \citenamefont {Reddy}, \citenamefont {Cournapeau}, \citenamefont
  {Burovski}, \citenamefont {Peterson}, \citenamefont {Weckesser},
  \citenamefont {Bright} \emph {et~al.}}]{virtanen2020scipy}%
  \BibitemOpen
  \bibfield  {author} {\bibinfo {author} {\bibfnamefont {P.}~\bibnamefont
  {Virtanen}}, \bibinfo {author} {\bibfnamefont {R.}~\bibnamefont {Gommers}},
  \bibinfo {author} {\bibfnamefont {T.~E.}\ \bibnamefont {Oliphant}}, \bibinfo
  {author} {\bibfnamefont {M.}~\bibnamefont {Haberland}}, \bibinfo {author}
  {\bibfnamefont {T.}~\bibnamefont {Reddy}}, \bibinfo {author} {\bibfnamefont
  {D.}~\bibnamefont {Cournapeau}}, \bibinfo {author} {\bibfnamefont
  {E.}~\bibnamefont {Burovski}}, \bibinfo {author} {\bibfnamefont
  {P.}~\bibnamefont {Peterson}}, \bibinfo {author} {\bibfnamefont
  {W.}~\bibnamefont {Weckesser}}, \bibinfo {author} {\bibfnamefont
  {J.}~\bibnamefont {Bright}}, \emph {et~al.},\ }\bibfield  {title} {\bibinfo
  {title} {Scipy 1.0: fundamental algorithms for scientific computing in
  python},\ }\href@noop {} {\bibfield  {journal} {\bibinfo  {journal} {Nat.
  Methods}\ }\textbf {\bibinfo {volume} {17}},\ \bibinfo {pages} {261}
  (\bibinfo {year} {2020})}\BibitemShut {NoStop}%
\bibitem [{\citenamefont {Streltsov}\ \emph {et~al.}(2011)\citenamefont
  {Streltsov}, \citenamefont {Kampermann},\ and\ \citenamefont
  {Bruss}}]{streltsov2011simple}%
  \BibitemOpen
  \bibfield  {author} {\bibinfo {author} {\bibfnamefont {A.}~\bibnamefont
  {Streltsov}}, \bibinfo {author} {\bibfnamefont {H.}~\bibnamefont
  {Kampermann}},\ and\ \bibinfo {author} {\bibfnamefont {D.}~\bibnamefont
  {Bruss}},\ }\bibfield  {title} {\bibinfo {title} {Simple algorithm for
  computing the geometric measure of entanglement},\ }\href@noop {} {\bibfield
  {journal} {\bibinfo  {journal} {Physical Review A}\ }\textbf {\bibinfo
  {volume} {84}},\ \bibinfo {pages} {022323} (\bibinfo {year}
  {2011})}\BibitemShut {NoStop}%
\bibitem [{\citenamefont {DiVincenzo}\ \emph {et~al.}(2003)\citenamefont
  {DiVincenzo}, \citenamefont {Mor}, \citenamefont {Shor}, \citenamefont
  {Smolin},\ and\ \citenamefont {Terhal}}]{divincenzo2003unextendible}%
  \BibitemOpen
  \bibfield  {author} {\bibinfo {author} {\bibfnamefont {D.~P.}\ \bibnamefont
  {DiVincenzo}}, \bibinfo {author} {\bibfnamefont {T.}~\bibnamefont {Mor}},
  \bibinfo {author} {\bibfnamefont {P.~W.}\ \bibnamefont {Shor}}, \bibinfo
  {author} {\bibfnamefont {J.~A.}\ \bibnamefont {Smolin}},\ and\ \bibinfo
  {author} {\bibfnamefont {B.~M.}\ \bibnamefont {Terhal}},\ }\bibfield  {title}
  {\bibinfo {title} {Unextendible product bases, uncompletable product bases
  and bound entanglement},\ }\href {https://doi.org/10.1007/s00220-003-0877-6}
  {\bibfield  {journal} {\bibinfo  {journal} {Commun. Math. Phys.}\ }\textbf
  {\bibinfo {volume} {238}},\ \bibinfo {pages} {379} (\bibinfo {year}
  {2003})}\BibitemShut {NoStop}%
\bibitem [{\citenamefont {Wei}\ and\ \citenamefont
  {Severini}(2010)}]{wei2010matrix}%
  \BibitemOpen
  \bibfield  {author} {\bibinfo {author} {\bibfnamefont {T.-C.}\ \bibnamefont
  {Wei}}\ and\ \bibinfo {author} {\bibfnamefont {S.}~\bibnamefont {Severini}},\
  }\bibfield  {title} {\bibinfo {title} {Matrix permanent and quantum
  entanglement of permutation invariant states},\ }\bibfield  {journal}
  {\bibinfo  {journal} {J. Math. Phys.}\ }\textbf {\bibinfo {volume} {51}},\
  \href {https://doi.org/10.1063/1.3464263} {10.1063/1.3464263} (\bibinfo
  {year} {2010})\BibitemShut {NoStop}%
\bibitem [{\citenamefont {Gharahi}\ and\ \citenamefont
  {Mancini}(2021)}]{gharahi2021algebraic}%
  \BibitemOpen
  \bibfield  {author} {\bibinfo {author} {\bibfnamefont {M.}~\bibnamefont
  {Gharahi}}\ and\ \bibinfo {author} {\bibfnamefont {S.}~\bibnamefont
  {Mancini}},\ }\bibfield  {title} {\bibinfo {title} {Algebraic-geometric
  characterization of tripartite entanglement},\ }\href
  {https://doi.org/10.1103/physreva.104.042402} {\bibfield  {journal} {\bibinfo
   {journal} {Phys. Rev. A}\ }\textbf {\bibinfo {volume} {104}},\ \bibinfo
  {pages} {042402} (\bibinfo {year} {2021})}\BibitemShut {NoStop}%
\bibitem [{\citenamefont {H{\"u}bener}\ \emph {et~al.}(2009)\citenamefont
  {H{\"u}bener}, \citenamefont {Kleinmann}, \citenamefont {Wei}, \citenamefont
  {Gonz{\'a}lez-Guill{\'e}n},\ and\ \citenamefont
  {G{\"u}hne}}]{hubener2009geometric}%
  \BibitemOpen
  \bibfield  {author} {\bibinfo {author} {\bibfnamefont {R.}~\bibnamefont
  {H{\"u}bener}}, \bibinfo {author} {\bibfnamefont {M.}~\bibnamefont
  {Kleinmann}}, \bibinfo {author} {\bibfnamefont {T.-C.}\ \bibnamefont {Wei}},
  \bibinfo {author} {\bibfnamefont {C.}~\bibnamefont
  {Gonz{\'a}lez-Guill{\'e}n}},\ and\ \bibinfo {author} {\bibfnamefont
  {O.}~\bibnamefont {G{\"u}hne}},\ }\bibfield  {title} {\bibinfo {title}
  {Geometric measure of entanglement for symmetric states},\ }\href
  {https://doi.org/10.1103/physreva.80.032324} {\bibfield  {journal} {\bibinfo
  {journal} {Phys. Rev. A}\ }\textbf {\bibinfo {volume} {80}},\ \bibinfo
  {pages} {032324} (\bibinfo {year} {2009})}\BibitemShut {NoStop}%
\bibitem [{\citenamefont {Chitambar}\ \emph {et~al.}(2008)\citenamefont
  {Chitambar}, \citenamefont {Duan},\ and\ \citenamefont
  {Shi}}]{chitambar2008tripartite}%
  \BibitemOpen
  \bibfield  {author} {\bibinfo {author} {\bibfnamefont {E.}~\bibnamefont
  {Chitambar}}, \bibinfo {author} {\bibfnamefont {R.}~\bibnamefont {Duan}},\
  and\ \bibinfo {author} {\bibfnamefont {Y.}~\bibnamefont {Shi}},\ }\bibfield
  {title} {\bibinfo {title} {Tripartite entanglement transformations and tensor
  rank},\ }\href {https://doi.org/10.1103/physrevlett.101.140502} {\bibfield
  {journal} {\bibinfo  {journal} {Phys. Rev. Lett.}\ }\textbf {\bibinfo
  {volume} {101}},\ \bibinfo {pages} {140502} (\bibinfo {year}
  {2008})}\BibitemShut {NoStop}%
\bibitem [{\citenamefont {Christandl}\ \emph {et~al.}(2023)\citenamefont
  {Christandl}, \citenamefont {Lysikov}, \citenamefont {Steffan}, \citenamefont
  {Werner},\ and\ \citenamefont {Witteveen}}]{christandl2023resource}%
  \BibitemOpen
  \bibfield  {author} {\bibinfo {author} {\bibfnamefont {M.}~\bibnamefont
  {Christandl}}, \bibinfo {author} {\bibfnamefont {V.}~\bibnamefont {Lysikov}},
  \bibinfo {author} {\bibfnamefont {V.}~\bibnamefont {Steffan}}, \bibinfo
  {author} {\bibfnamefont {A.~H.}\ \bibnamefont {Werner}},\ and\ \bibinfo
  {author} {\bibfnamefont {F.}~\bibnamefont {Witteveen}},\ }\bibfield  {title}
  {\bibinfo {title} {The resource theory of tensor networks},\ }\href@noop {}
  {\bibfield  {journal} {\bibinfo  {journal} {arXiv preprint arXiv:2307.07394}\
  } (\bibinfo {year} {2023})}\BibitemShut {NoStop}%
\bibitem [{\citenamefont {Landsberg}(2006)}]{landsberg2006border}%
  \BibitemOpen
  \bibfield  {author} {\bibinfo {author} {\bibfnamefont {J.}~\bibnamefont
  {Landsberg}},\ }\bibfield  {title} {\bibinfo {title} {The border rank of the
  multiplication of 2$\times$ 2 matrices is seven},\ }\href
  {https://doi.org/10.1090/s0894-0347-05-00506-0} {\bibfield  {journal}
  {\bibinfo  {journal} {J. Am. Math. Soc.}\ }\textbf {\bibinfo {volume} {19}},\
  \bibinfo {pages} {447} (\bibinfo {year} {2006})}\BibitemShut {NoStop}%
\bibitem [{\citenamefont {Bhat}(2006)}]{bhat2006completely}%
  \BibitemOpen
  \bibfield  {author} {\bibinfo {author} {\bibfnamefont {B.~R.}\ \bibnamefont
  {Bhat}},\ }\bibfield  {title} {\bibinfo {title} {A completely entangled
  subspace of maximal dimension},\ }\href
  {https://doi.org/10.1142/s0219749906001797} {\bibfield  {journal} {\bibinfo
  {journal} {Int. J. Quantum Inf.}\ }\textbf {\bibinfo {volume} {4}},\ \bibinfo
  {pages} {325} (\bibinfo {year} {2006})}\BibitemShut {NoStop}%
\bibitem [{\citenamefont {Zhang}\ and\ \citenamefont
  {Zhu}(2024)}]{zhang_2024_12166896}%
  \BibitemOpen
  \bibfield  {author} {\bibinfo {author} {\bibfnamefont {C.}~\bibnamefont
  {Zhang}}\ and\ \bibinfo {author} {\bibfnamefont {X.}~\bibnamefont {Zhu}},\
  }\href {https://doi.org/10.5281/zenodo.12166896} {\bibinfo {title}
  {numqi/entangled-subspace}} (\bibinfo {year} {2024})\BibitemShut {NoStop}%
\end{thebibliography}%
\end{document}